\documentclass[fleqn,usenatbib]{mnras}

\usepackage{newtxtext,newtxmath}

\usepackage[T1]{fontenc}
\usepackage{ae,aecompl}

\usepackage{graphicx}	
\usepackage{amsmath}	
\usepackage{amssymb}	
\usepackage{xfrac}
\usepackage{gensymb}

\newtheorem{theorem}{Theorem}
\newtheorem{definition}{Definition}
\newtheorem{proposition}{Proposition}
\newtheorem{corollary}{Corollary}

\renewcommand{\leq}{\leqslant}
\renewcommand{\geq}{\geqslant}

\newcommand{\bs}{\boldsymbol}

\newcommand{\ie}{i.e., }
\newcommand{\eg}{e.g., }

\title[STIM map: detection map for exoplanets imaging]{STIM map: detection map for exoplanets imaging\\ beyond asymptotic Gaussian residual speckle noise} 

\author[B. Pairet et al.]{
Beno\^it Pairet,$^{1}$\thanks{Email: benoit.pairet@uclouvain.be}\thanks{BP, OA and LJ are funded by the Belgian F.R.S.-FNRS. Part of this study is funded by the project AlterSense (MIS-FNRS)}
Faustine Cantalloube$^2$, Carlos A. Gomez Gonzalez$^3$\thanks{CGG and OA acknowledge funding from the European Research Council under the European Union's Seventh Framework Programme (ERC Grant Agreement No. 337569) and from the French Community of Belgium through an ARC grant for Concerted Research Action.},  \newauthor  Olivier Absil$^4$\textcolor{blue}{\footnotemark[2]\footnotemark[3]}, Laurent Jacques$^1$\textcolor{blue}{\footnotemark[2]}\\
$^1$ ISPGroup, ELEN/ICTEAM, UCLouvain, Belgium\\
$^2$ Max Planck Institute for Astronomy, Germany\\
$^3$ Universit\'e Grenoble Alpes, IPAG, F-38000, Grenoble, France \\
$^4$ Space sciences, Technologies, and Astrophysics Research (STAR) Institute, University of Li\`{e}ge, Belgium 
}

\date{Accepted XXX. Received YYY; in original form ZZZ}

\pubyear{2019}

\begin{document}
\label{firstpage}
\pagerange{\pageref{firstpage}--\pageref{lastpage}}
\maketitle

\begin{abstract}
Direct imaging of exoplanets is a challenging task as it requires to reach a high contrast at very close separation to the star.
    Today, the main limitation in the high-contrast images is the quasi-static speckles that are created by residual instrumental aberrations. They have the same angular size as planetary companions and are often brighter, hence hindering our capability to detect exoplanets.
    Dedicated observation strategies and signal processing techniques are necessary to disentangle these speckles from planetary signals. 
    The output of these methods is a detection map in which the value of each pixel is related to a probability of presence of a planetary signal.
        The detection map found in the literature relies on the assumption that the residual noise is Gaussian. However, this is known to lead to higher false positive rates, especially close to the star. In this paper, we re-visit the notion of detection map by analyzing the speckle noise distribution, namely the Modified Rician distribution. We use non-asymptotic analysis of the sum of random variables to show that the tail of the distribution of the residual noise decays as an exponential distribution, hence explaining the high false detection rate obtained with the Gaussian assumption.
From this analysis, we introduce a novel time domain detection map and we demonstrate its capabilities and the relevance of our approach through experiments on real data. We also provide an empirical rule to determine detection threshold providing a good trade off between true positive and false positive rates for exoplanet detection.
\end{abstract}



\section{Introduction}

In the field of exoplanet study, high contrast imaging (HCI) provides valuable information to study planetary systems properties since it gives access to the spectral features of the exoplanet's atmosphere~\citep{Crossfield2015atmosphere,konopacky2013detection}, its mass determination via orbital follow-up~\citep{pueyo2015reconnaissance,Bonnefoy2014astrometry}, and the study of its interactions with its environment such as other planets or circumstellar disks~\citep{Espaillat2014ppdisk,Hughes2018debrisdisk}. This information can constrain planetary system formation models and improve our understanding of the nature of exoplanets~\citep[see][for a review]{Bowler2018reviewEP}.

Only a few tens of exoplanets have been directly detected around the hundreds of stars observed within surveys led during the last decade~\citep{Chauvin2015survey}. This low number of detections tells us that either the types of planets accessible through direct imaging are indeed rare or that our sample is strongly biased by our technical limitations. The main challenge of exoplanet imaging is that exoplanets are faint objects located in close vicinity to their host star that is much brighter. Emitted light from young Jupiter-like planets are typically $10^{-6}$-$10^{-4}$ fainter than their host star in the near infrared, where the contrast is favorable, and the typical projected separation between the planet and its host star is of $0.1$ arcsec. 

To reach such an angular resolution, 10-m class ground based telescopes are used in combination with adaptive optics (AO) systems which correct for the resolution loss induced by the atmospheric turbulence~\citep{Guyon2005ao}. Coronagraph devices are then used to increase the dynamic range by removing the coherent part of the starlight which is hiding the faint circumstellar signals~\citep{Guyon2006coro}. With dedicated instruments such as VLT/SPHERE~\citep{Beuzit2008sphere}, Gemini/GPI~\citep{Macintosh2008gpi}, KecK/NIRC2~\citep{McLean2000nirc2}, Subaru/SCeXAO~\citep{Jovanovic2015scexao} or LBTI/LMIRCam~\citep{Hinz2016lmircam}, a typical contrast of $10^{-4}$ is obtained at a separation of 0.5 arcsec. However, residual aberrations induce the presence of speckles in the images, which are of the same angular size as a point source and are often brighter than the exoplanets signal. These speckles are quasi-statics and hence cannot be calibrated nor averaged through longer exposures. Post-processing techniques are then applied to disentangle the planetary signals from these starlight residuals and thus reach a contrast of down to $10^{-6}$ at 0.5 arcsec~\citep{mawet2012review}.

To disentangle signals, post-processing methods require diversity within the data. The diversity is obtained through specific observations strategies (such as pupil tracking, dual band imaging or dual polarization imaging). The most widely used methods today rely on  Angular Differential Imaging ~\citep[ADI][]{marois2006angular} which makes use of pupil tracking mode observations. This mode keeps the speckle field almost constant during the observation while any circumstellar signal rotates with a deterministic velocity given by the parallactic angles. Most methods today consist in  empirically estimating the speckle field, then subtracting it from each frame of the image cube, and then combining the frames to form the so-called \emph{processed frame}. In this processed frame the residual speckles add up incoherently and thus average to small values whereas planetary signals are aligned and average to the actual values of the intensity of the planet.
 
Once this processed frame is computed, one has to perform the detection, which is usually visually performed by the user. A robust detection requires the use of statistical tests from which a detection map is built. In this context, a good knowledge of the underlying distribution of the residual noise is crucial. As discussed further below, current estimation techniques rely on asymptotic analysis of random variables that predicts a Gaussian distribution. As it is known that a Gaussian based detection leads to high false positive rate~\citep{marois2008confidence}, it has been proposed to modify the detection procedure to take into account the deviation from Gaussianity. However, to the best of the knowledge of the authors, no study has yet proposed a method to measure how far off the residuals are from a Gaussian distribution.

In the present work, we argue that the tail of the distribution is the important feature as it will decide the outcome of the statistical test. We empirically and theoretically discuss the decay of the tail of the distribution of the residual noise after post-processing. This analysis is non-asymptotic and the resulting estimate of the residual noise level depends on the number of frames. Then we introduce and study a new detection map that is built in light of our results. Although our detection map is designed for an ADI sequence of images, we stress that the non-asymptotic analysis of the speckle noise distribution does not depend on the observation strategy and can be applied to any images where speckle noise is present.

 \paragraph*{Contributions:} This work brings several contributions, summarized below. 

First, we study the tail decay of the residual noise distribution on the processed frame for several datasets. We show numerically that in each case the (empirical) quantiles of the noise distribution are closer to the quantiles of a Laplacian distribution than to those of a Gaussian distribution. This indicates that the residual noise exhibits the exponential tail decay of a Laplacian noise.

Second, we prove theoretically that the MR distribution is actually sub-exponential \citep{vershynin2010introduction}, \ie it belongs to a class of distributions --- including the Laplacian distribution --- whose tails decay exponentially. This allows us to use non-asymptotic statistical tools, based on measure concentration, to determine a meaningful estimate of the residual noise level on the processed frame in function of the number of frames. Moreover, we study the sensitivity of our analysis with respect to the fraction of frames that can be considered as statistically independent. In particular, we quantify how the temporal dependence between frames slows down the tail decay of the residual noise. We note that all these observations could not be reached with classical asymptotic analyses, \eg relying on the central limit theorem (CLT). 
 
Third, we leverage this non-asymptotic analysis to introduce a novel detection map, the standardized trajectory intensity mean map (or STIM map). Our theoretical analysis explains the distribution of pixel intensities in this new map, for which exoplanets are associated with clear outliers in a hypothesis testing context. Comparatively to the SNR$_t$ map \citep{mawet2014fundamental}, we observe that the STIM map distribution is also more concentrated where there is no exoplanet signal; exoplanet detection is thus made more stable with a single thresholding procedure. 

Finally, by establishing an automatic estimation of a reliable detection threshold, we demonstrate the capabilities of the STIM map through numerous experiments involving real datasets.

\paragraph*{Paper structure:} The rest of the paper is organized as follows. We present the state-of-the-art of the reduction techniques and of the detection procedure, our notations, and our conventions in Section~\ref{PremNotations}.In Section~\ref{sec:stateart}, after a brief presentation of the speckle noise distribution, we study the statistical properties of the residual speckle noise both empirically and theoretically, through a rigorous non-asymptotic statistical analysis. Then, we present how this analysis applies to the tail decay of the residual noise on the processed frame. We finally discuss the impact of non independence of the residual speckle noise. In Section~\ref{sec:stateart-detectionmap}, we review the current detection map procedure and we use our previous results to introduce and justify our new detection map, the STIM map. In Section~\ref{sec:simul}, we demonstrate the efficacy of our approach by computing the STIM map for several on-sky data from different instruments. We finally conclude and give perspectives for further applications of this work in Section~\ref{sec:conclu}. The information about the datasets used throughout the paper and the mathematical developments can be found in the appendices.

\section{Framework and Preliminaries}
\label{PremNotations}

\paragraph*{Current ADI-based post-processing techniques:} An ADI dataset is a volume with $T$ images (or frames) of size $n\times n$ that is reshaped into $\mathbb R^{T\times N}$ matrices with $N = n^2$. In the same fashion, $n\times n$ images are depicted as $ \mathbb R^{N}$ vectors. We denote matrices by capital bold symbols, \eg $\bs Y$, and vectors by lowercase bold symbols, \eg $\bs f$. 

\begin{figure}
  \centering  {\includegraphics[width=0.4\textwidth]{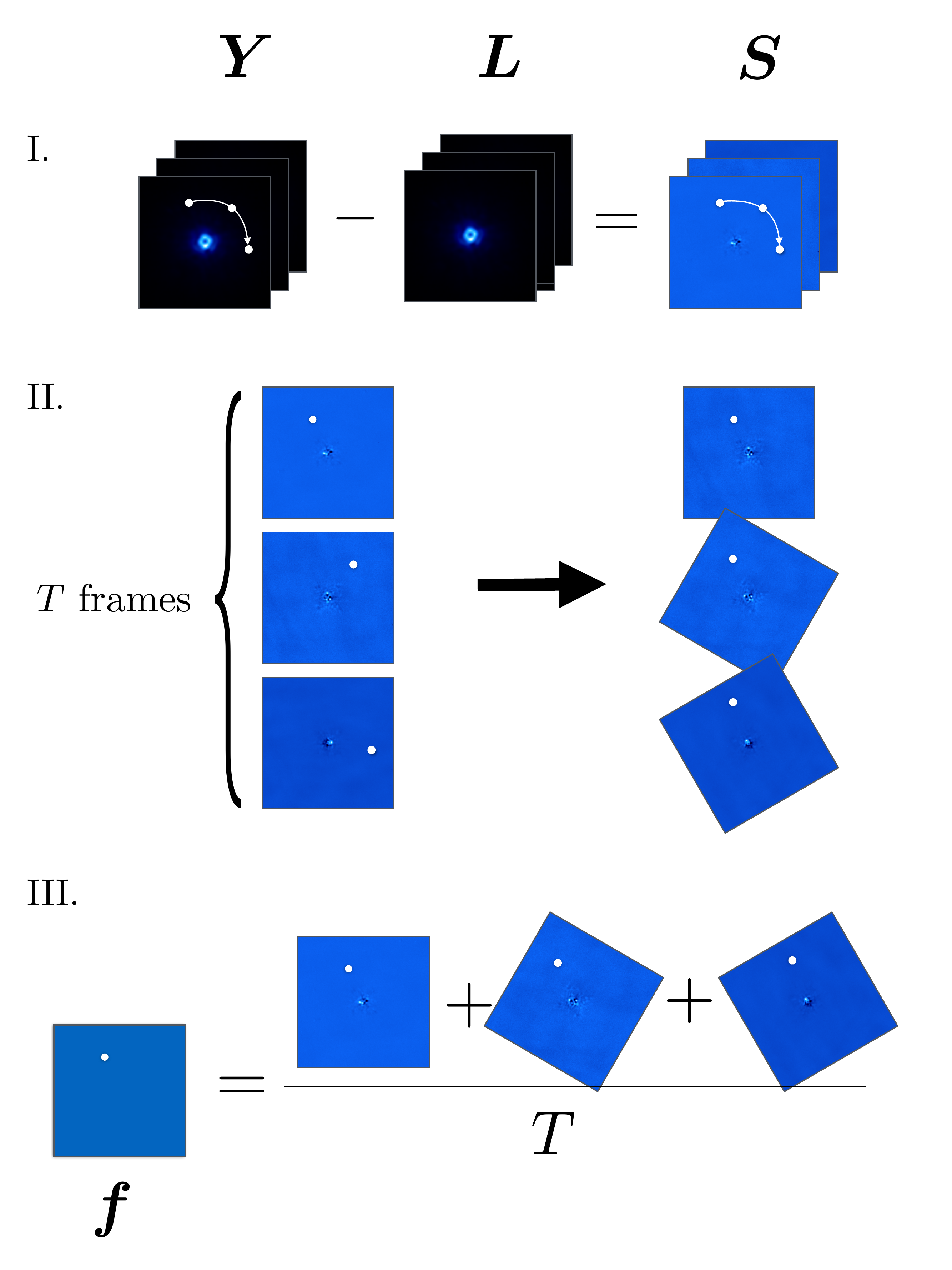}}
  \vspace{-2mm}\\
\caption{
Schematic representation of the ADI method. $\bs Y$ is the data cube, $\bs L$ is the model PSF, $\bs S$ the subtracted data cube and T the number of frames. $\bs f$ is the processed frame.}
\label{fig:ADI}
\end{figure}

ADI-based post-processing methods can be summarized in three steps described in Figure~\ref{fig:ADI}: \textit{(i)}~a point-spread function (PSF) model\footnote{We note that, in high contrast imaging, the PSF has a slightly different meaning than in other fields such as signal processing.~In the context of HCI for exoplanet detection, the PSF refers to the response of the whole observation system, from the star to the detector, going through the atmosphere, the telescope pupil, the adaptive optics, the coronagraph and any optical device in the light path. Hence here, the PSF model refers to the approximation of the whole starlight signal, accounting for the speckle field. Planetary signals have a different response because the coronagraph is designed to act mainly (if not only) on on-axis signals.} 
$\bs L$, containing only the host star signal with neither planetary companions nor circumstellar disks signals, is estimated empirically from the data cube $\bs Y$, \ie $\bs L = \Theta(\bs Y)$ for some function $\Theta$ discussed below, \textit{(ii)}~this PSF model is subtracted from the cube to form the volume $\bs S = \bs Y - \bs L$ and \textit{(iii)}~the individual frames of $\bs S$ are aligned to a common direction for the potential companions and collapsed into a processed frame $\bs f$. The temporal correlation of $\bs S$ is known to be lower than that of the initial volume, resulting in residual speckles being considerably less correlated. Thus in step \textit{(iii)}, the residual speckles average to a mean close to zero while the planetary signals add-up. In practice, part of the planetary signal is present in the model PSF $\bs L$ and the intensity of the planetary signals extracted from the resulting processed frame $\bs f$ is an underestimate of their true intensity.

The processed frame $\bs f$ computed in step \emph{(iii)} of the ADI-based postprocessing is often computed as the pixelwise median of the aligned frames of $\bs S$. In order to ease the theoretical analysis, we use in this study the mean instead of the median. Moreover, our methods are conveniently described according to the following mathematical conventions. When referring to a certain pixel in an image, we use a single index $g$ instead of a tuple $(i,j)$ such that an element of the processed frame is depicted as $f_g$. Instead of defining an aligned volume as in step \textit{(iii)} above, we first collect in a vector $\bs s^{[g]} = (s_1^{[g]},\cdots, s_T^{[g]})^\top \in \mathbb R^T$ the values of the pixels of~$\bs S$ that are part of the trajectory induced by the parallactic angles starting from the first line of the matrix $\bs S$ at index~$g$. 
In this context, the $g^{\text{th}}$ element of the processed frame is the computed mean of the trajectory $g$ in the volume $\bs S$:
\begin{equation}
  f_g = \hat \mu_g \equiv \hat \mu(\bs s^{[g]}) = \frac{1}{T}\sum_{i = 1}^{T}   s^{[g]}_i.
  \label{processed_frame_mu} 
  \end{equation}
This description of the post-processing in terms of trajectories is also illustrated  on Figure~\ref{Illustration_SNR_vs_STIM} in the context of the detection maps.

A wide variety of methods exists to construct the model PSF $\bs L$ (that is to say different ways to define the function $\Theta$), such as c-ADI~\citep[classical ADI][using the median of the data cube]{marois2006angular}, LOCI~\citep[Locally Optimised Combination of Images][using a linear combination of patches of the images]{lafreniere2007new}, or PCA~\citep[Principal Component Analysis][using the first principal components of the data cube]{soummer2012detection,amara2012pynpoint}. Other methods based on ADI perform different steps to obtain the processed frame $\bs f$, such as LLSG~\citep[][that separates the volume in a low-rank part for the star PSF plus a sparse part for the planetary signal]{gonzalez2016low}. We refer to this class of method as the speckle subtraction methods.

There exists another class of post-processing techniques based on the inverse problem approach that perform a maximum likelihood estimation (equivalent to matched filtering under the Gaussian hypothesis) of the companion flux and position, producing directly a detection map. The techniques, pioneered by ANDROMEDA~\citep{cantalloube2015direct}, have been extended in different fashions, such as multi-spectral data~\citep[see for instance the FMMF technique, ][]{ruffio2017improving}.

These two classes of post-processing methods are complementary in exoplanets detection. They are often used together, as it is illustrated, for instance, in~\cite{DelormeInDepth}.

The aim of the detection map that we introduce in Section~\ref{sec:STIMdetectionmap} is to provide a more robust detection procedure for speckle subtraction methods. For this reason, we do not consider the second class of methods in the present paper. However, it is worth mentioning that our theoretical findings have implications in the maximum likelihood estimation on which they rely. This is briefly discussed in the conclusion.

\paragraph*{Detection procedure:} We want to test the absence or the presence of a planet for each pixel $g$ on the processed frame $\bs f$. In this context, we define on the trajectory supported by $g$ the null-hypothesis $H_0$ as the absence of a planet and the research hypothesis $H_1$ as the presence of a planet. Mathematically, 
\begin{align}
 H_0:& \quad f_{g} = I_{\text{noise}}, \\
 H_1:& \quad f_{g} = I_{\text{planet}} + I_{\text{noise}},
\end{align}
where $I_{\text{planet}}$ is the intensity of an hypothetical planet on location $g$ and $I_{\text{noise}}$ is a random variable describing the residual noise at this location.

Nothing is known \emph{a priori} about the value of $I_{\text{planet}}$, hence we accept $H_1$ by rejecting $H_0$. Given an observed intensity $I$, how likely is it that the null-hypothesis accounts for this observation? In other words, what is the probability that the random variable $I_{\text{noise}}$ takes a value equal or greater than $I$. Or mathematically, how large is $P \left( f_g \geq  I | H_0 \right) = P \left( I_{noise} \geq  I \right) $. If it is unlikely that the residual noise explains the observed values, then it is likely that something else, such as an off-axis signal, explains it. 

In this context, it is thus important to have a realistic estimate of the distribution of the residual noise. Then one can select a detection threshold in order to have a fixed confidence level, \ie a fixed probability, for a planetary signal to be detected.
 
Because the usual data reducing techniques have a whitening effect on the residuals, justifying the independence of the random variables summed together into the processed frame and the number of frames being typically large, the central limit theorem (CLT) is generally invoked to state that the residual noise in the processed frame follows a Gaussian distribution~\citep{marois2008confidence, mawet2014fundamental}. Hence, under the null hypothesis, pixels values are assumed to be drawn from a Gaussian distribution with zero mean and standard deviation $\sigma$. The probability of observing a value of $5\sigma$ is below $3\times 10^{-7}$. Hence, rejecting $H_0$ when $I_g > 5\sigma$ yields a confidence level of $1-3\times 10^{-7}$ under a Gaussian assumption. However, it is known that the Gaussian assumption leads to high false positive rates~\citep{marois2008confidence, mawet2014fundamental}. We briefly summarize how this non-Gaussianity is currently accounted for in Section~\ref{sec:stateoftheart_detectionmap}, before introducing the proposed detection procedure based on our theoretical analysis.

In the context of direct imaging, the outcome of the hypothesis testing is a detection map that assigns a value to each pixel in the processed frame. The larger this value, the less likely the null-hypothesis is verified, hence the more likely a planetary signal is present at that location.

\paragraph*{Planet free datasets:} In what follows, we will study the distribution of the residual speckle noise in the processed frame. The presence of a planet disturbs the tail distribution as planets are precisely detected as outliers in the processed frame. Hence the study of the tail decay requires planet-free datasets. An option is to remove all known planetary signals with, for instance, the negative companion injection method~\citep[NEGFC]{Marois2010aNEGFC}. However, this method does not guarantee that the totality of the planetary signal is removed as it cannot be used for extended sources or faint planetary signals that are not previously detected. 

For these reasons, we decided to use another method to significantly reduce the influence of potential exoplanets or other on-sky signals in the dataset. We consider the trajectory groups $\bar g \in \bar{\mathcal G}$ that we obtain using the opposite values of the parallactic angles \citep{marois2008confidence}. This way, we obtain a similar temporal dependence of the residual speckles but circumstellar signals will be all averaged to negligible values and have larger standard deviation. Processed frames obtained with this method are referred to as \emph{opposite angles processed frames} in the text. We discuss the applicability of this method in Section~\ref{sec:simul}. 

In the following, we will use three datasets described in~Appendix~\ref{app-data_set}: $\beta$-Pic using the VLT-NACO instrument, HD 206893 and 51 Eri taken with the VLT/SPHERE-IRDIS instrument. Each target hosts a planetary signal.

\section{Residual speckles statistics}
\label{sec:stateart}

In this section, we first present the statistics followed by the speckle noise. Then we empirically show that the distribution of the residual noise on the processed frame exhibits a slower decay than expected with a Gaussian distribution. After that, we introduce the concept of sub-exponentiality and show that the MR distribution is sub-exponential. We use this newly demonstrated property of the speckle noise to characterize the residual noise on the processed frame using non-asymptotic statistics. We end this section by an analysis of the impact of the non-independence of the residuals on $\bs S$.

\subsection{Speckle noise statistics} 

The mean intensity $I$ for an AO-corrected long exposure can be modeled as the sum of the static coherent point spread function (only due to the diffraction by the telescope aperture) $I_c$ and a random speckle noise intensity $I_s$. It has been shown that the total intensity $I$ follows a Modified Rician (MR) distribution:
\begin{equation}
p_{\text{MR}}(I , I_c,I_s) = \frac{1}{I_s}\exp \left(-\frac{I+I_c}{I_s}\right) \mathcal{I}_0 \left(\frac{2\sqrt{II_c}}{I_s}\right),
\label{MR_pdf}
\end{equation}
where $\mathcal{I}_0$ is the modified Bessel function of the first kind. This equation, first derived for laser~\citep{goodman1975statistical}, was adapted to high contrast imaging for exoplanet detection~\citep[][and references therein]{fitzgerald2006speckle, soummer2007speckle,marois2008confidence}.

The expectation and variance of $I$ are given by~\citep{soummer2007speckle}:
\begin{align}
\mathbb E I &= I_s + I_c,\\
\sigma^2_I &= I^2_s + 2I_sI_c.
\end{align}

In the following, we will consider that in high flux regime the other sources of noise (photon and detector noise) are negligible, especially close to the star, and hence $\sigma_{\text{total}} = \sigma_I$. 

The parameters $I_c$ and $I_s$ are not constant throughout the field of view, however they can be consider as constant at a given radius, \ie at constant separation from the star. Hence the the mean and standard deviation of the speckle is a function of the radial distance from the star. 

When $I_c$ equals zero, the intensity distribution turns into a pure speckle exponential statistics. Note that when using a coronagraph, the unaberrated term $I_c$ tends toward 0 since the coronagraph is designed to remove the static diffraction pattern. Moreover, thanks to the post-processing, the PSF model $\bs L$ which is subtracted to the data cube includes the static features and hence $I_s \gg I_c$. 

Another important aspect of the speckle noise is the different timescales appearing in the speckle formation~\citep{hinkley2007temporal}. A thorough analysis of the impact these different timescales is beyond the scope of the present paper. However, we argue~\citep{soummer2007speckle} that this effect can reasonably be modeled by changing $I_s$ to $I_{s_1} + I_{s_2}$ where $I_{s_1}$ and $I_{s_2}$ are the random speckle noise intensity of timescales $\tau_1$ and $\tau_2$ respectively. This substitution has no impact in the scope of the present work, hence, for simplicity and without loss of generality, we consider the distribution displayed in Equation~\eqref{MR_pdf} for the speckles.

\subsection{Tail-decay of the residual noise on the processed frame}
\label{sec:MRsubexp}

\begin{figure*}
  \centering 
      {\includegraphics[width=0.3\textwidth]{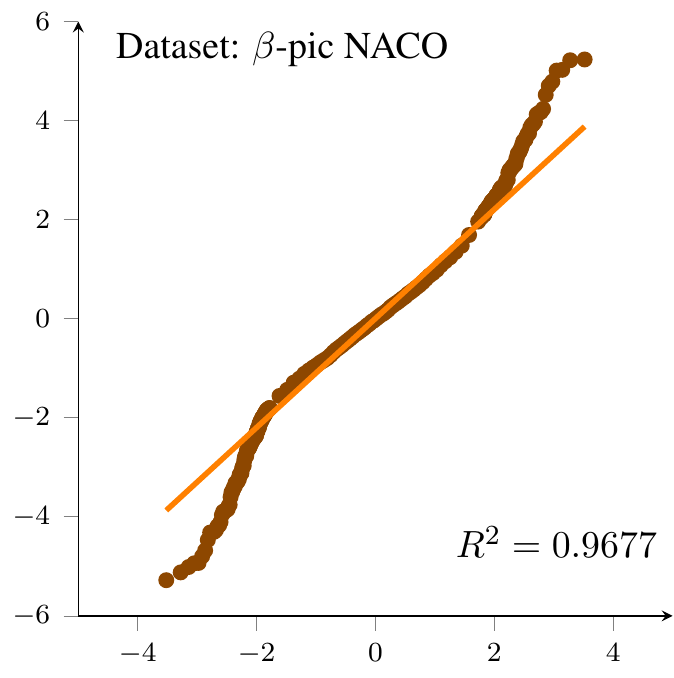}}
   \hspace{0.2cm}
        {\includegraphics[width=0.3\textwidth]{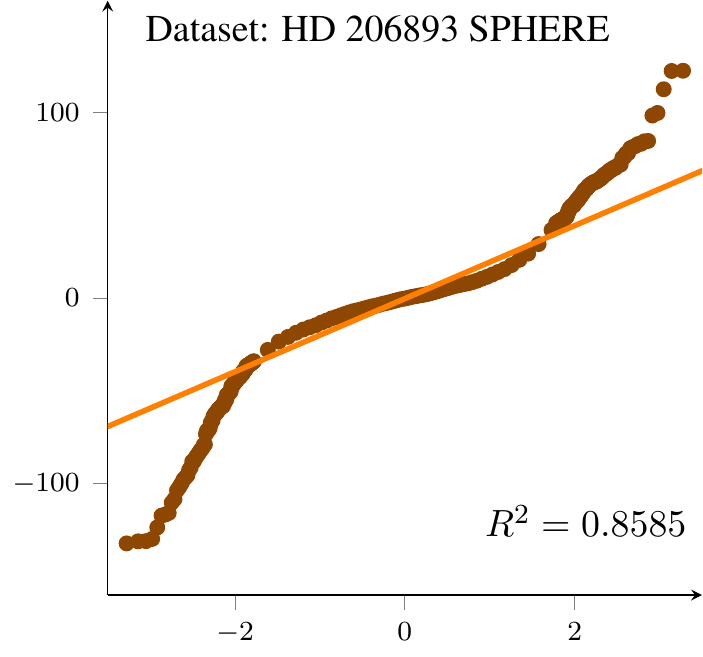}} 
   \hspace{0.2cm}
{\includegraphics[width=0.3\textwidth]{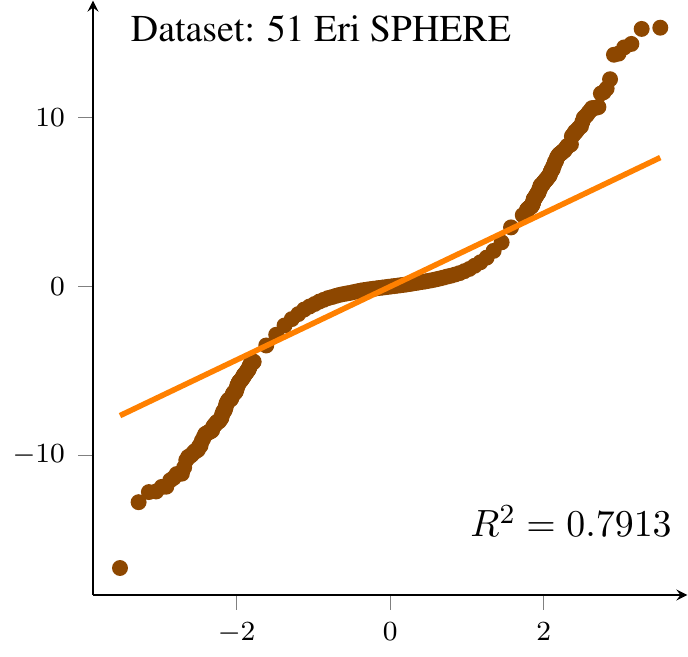}} \\
   \vspace{0.2cm}
        {\includegraphics[width=0.3\textwidth]{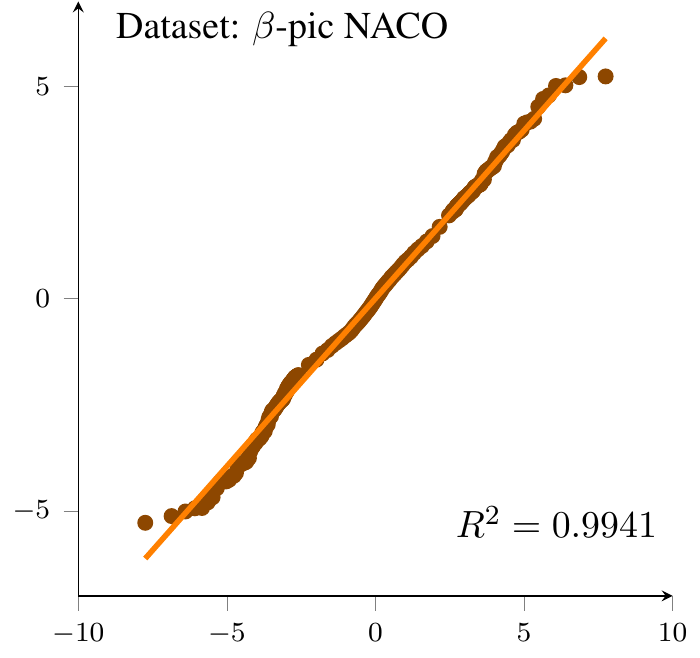}} 
   \hspace{0.2cm}
        {\includegraphics[width=0.3\textwidth]{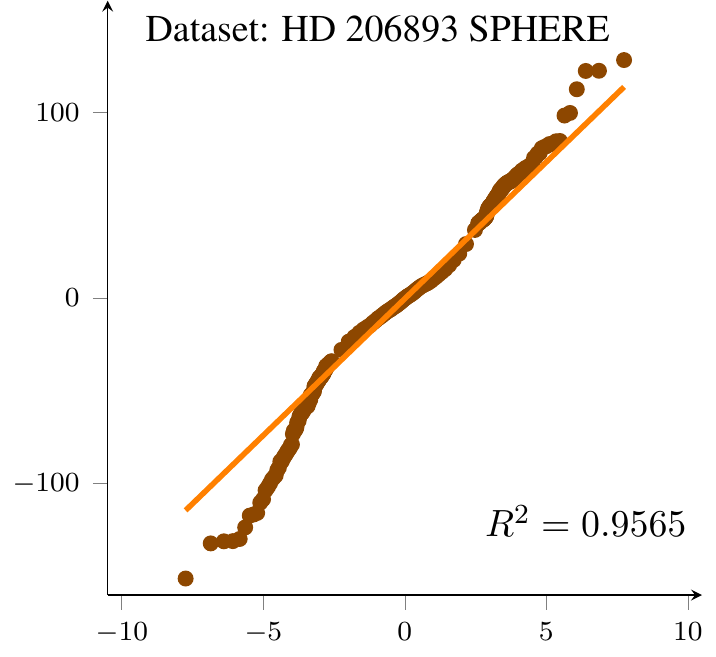}} 
   \hspace{0.2cm}
        {\includegraphics[width=0.3\textwidth]{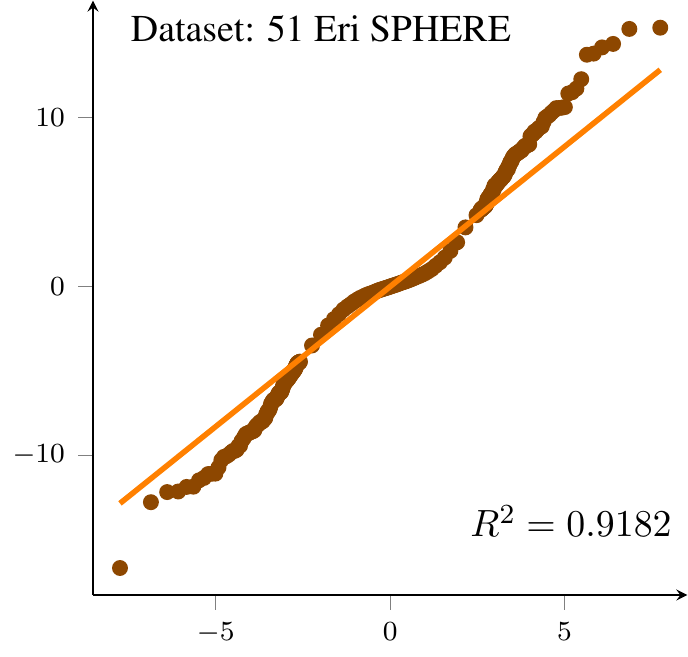}} 
  \caption{Q-Q plots for the three datasets. The sample is drawn from the center circle of radius of $8\lambda/D$ of the opposite angles processed frame, using respectively 10, 17 and 5 principal components. Top, we compare the distribution of the pixel values to the Gaussian distribution. We can see that the fit is accurate for the center of the distribution but becomes increasingly bad for the end of the tails. Bottom, we make the comparison with the Laplace distribution. The corresponding Q-Q plot being closer to a straight line than under the Gaussian assumption. We display $R^2$, the coefficient of determination, \ie the Pearson correlation between the paired of quantiles. The closer $R^2$ is to $1$, the better the fit is, \ie if the two distributions are linearly related, then $R^2=1$. Its values confirms here that the Laplacian fit is better than the Gaussian fit.} 
  \label{fig:qq_plot}
\end{figure*}

We use a statistical visualization tool, namely a quantile-quantile plot (Q-Q plot), to empirically show that, under the null-hypothesis, the tail decay of pixel intensities on the processed frame is better explained by a Laplacian distribution than by a Gaussian distribution. This empirical observation is, later in this paper, supported by a theoretical analysis of the MR distribution.

A Q-Q plot is a statistical tool to compare two distributions by plotting their respective quantiles against each other \citep[see, \eg][]{heiberger2004statistical}. It thus allows one to compare the empirical distribution of some data to the distribution they are assumed to follow. If the assumed distribution is correct, the resulting Q-Q plot will approximately be a straight line. When the distribution parameters (\eg mean and variance) are identical, the resulting straight line is the bisector. Deviations from the straight line indicate differences between the two distributions, such as  different skewness or kurtosis. In the context of exoplanet detection, we are particularly interested in the tail of the distribution, as this indicates when the $H_0$ hypothesis can be rejected. As we are only interested in the positive values, the information about the tail is extracted from the right side of the plot. If the points of the Q-Q plot are above the straight line in the right side of the plot, it indicates that the distribution on the $y$ axis (data) has a heavier tail than the distribution on the $x$ axis (test distribution). 

We use the opposite angles processed frames for all three datasets in order to significantly reduce the influence of any circumstellar signal which could bias the residual speckles distribution (as described at the end of Section~\ref{PremNotations}). To build the Q-Q plots, we arbitrarily select the pixels within the central annulus of radius $8\lambda/D$ in the opposite angles processed frame as our data sample. Note that we observe that the results are similar when performed with a different radius or annulus-wise.  

We compare the pixel intensities distribution of our sample to the Gaussian distribution in the Q-Q plot displayed on Figure~\ref{fig:qq_plot} (top). We see that the Gaussian fit is good for the first quantiles but becomes increasingly discrepant for higher quantiles. As a consequence, likelihood that residual noise has a large value is greater than what we can expect from a Gaussian random variable. In the framework of detection, it means that a given probability of presence under the Gaussian assumption will in reality yield a larger number of false alarms. 

Interestingly, we found that a Laplace distribution yields a better fit for the tail decay, as shown on Figure~\ref{fig:qq_plot} (bottom). The Laplace distribution follows a probability density function $f(x|\mu,b) = \frac{1}{2b} \exp\left(-\frac{\lvert x-\mu \rvert}{b}\right)$, where $\mu$ is the mean and $b$ refers to as the diversity, which is linked to the variance as $\sigma^2 = 2b^2$. For $x\geq \mu$, the cumulative distribution function (CDF) is given by \[F_x(x) = P(X \leq x) = 1 - \frac{1}{2} \exp(-\frac{x-\mu}{b}).\] 
In other words, this means that the probability that a Laplacian random variable is larger than the mean, $P(X - \mu \geq x) = \frac{1}{2} \exp(-\frac{x}{b})$ decreases exponentially. In comparison, for a Gaussian r.v.'s, we have $P(X -\mu \geq x) \leq  \exp(-\frac{x^2}{2\sigma^2})$ when $x\geq 0$ increases.\\

To ease the comparison between the Gaussian and the Laplacian plots, we also displayed on Figure~\ref{fig:qq_plot} the coefficient of determination $R^2$ that is the Pearson correlation between the paired sample quantiles. For two compared samples $X$ and $Y$, it is given by: \[R^2 = \left( \frac{\text{cov}(X,Y)}{\hat \sigma(X) \hat \sigma(Y)} \right)^2,\]
where $\text{cov}(X,Y)$ is the sample covariance between samples $X$ and $Y$. 
The closer $R^2$ is to 1, the closer the distribution $X$ is to the distribution $Y$. We can see that $R^2$ is always closer to one in the Laplacian probability plots than in the Gaussian probability plots, thus supporting the observed trends.

We emphasize here that the only information provided by Figure~\ref{fig:qq_plot} is that the tail decay is better explained by a Laplacian (\ie it has an exponential tail decay) but not that the actual distribution of the residual noise on the processed frame is Laplacian. A bound for the actual tail decay of the residual noise on the processed frame is derived in Section~\ref{sec:sub_exp_processed_frame}, based on the theoretical results presented in Section~\ref{subExp_MR_dist}.

\subsection{Sub-exponentiality of the MR distribution}
\label{subExp_MR_dist}

We prove now that an MR distribution has an exponential tail day, \ie it belongs to the class of \emph{sub-exponential} random variables \citep{vershynin2010introduction}. Consequently, a MR random variable $Z$ is such that $P(Z>z)$ decays like $\mathcal O(e^{-z})$ when the level $z$ increases, and not like the Gaussian tail decay $\mathcal O (e^{-z^2})$. As described below (see Theorem~\ref{th:subexp_decay}), this has a clear impact on the sum of $m$ independently and identically distributed (i.i.d.) MR random variables; this sum, seen as a random variable, has a tail that also decays exponentially when the level increases beyond a value only depending on the MR characteristics.

Mathematically, a sub-exponential random variable is such that~\citep{vershynin2010introduction},
\[
\textstyle P \left( \lvert X - \textstyle E(X) \rvert \geq t \right) \leq c_1 \exp(-c_2 t),
\]
for all $t\geq 0$, where $c_1, c_2> 0$ are two universal constants. Equivalently, sub-exponential random variables can be defined as follows~\citep{vershynin2010introduction}. 

\begin{definition}[Sub-exponential random variables] A random variable $X$ is called sub-exponential if its sub-exponential norm $\lVert X \rVert_{\psi_1}$, defined as
\begin{equation}
\lVert X \rVert_{\psi_1} := \sup_{p\geq 1}p^{-1} \left( \textstyle E \lvert X\rvert ^p\right)^{1/p},
\end{equation}
is bounded, \ie $\lVert X \rVert_{\psi_1} < +\infty$.
\label{def:subexp} 
\end{definition}

Interestingly, as proved in Appendix~\ref{ProofSubExpMRProp}, MR random variables are sub-exponential.
\begin{proposition}
Let $X \sim \text{MR}(\alpha,\beta)$ be a modified Riccian random variable with 
\[ \text{MR}(\alpha,\beta) \sim \frac{1}{\beta} \exp \left(-\frac{t+\alpha}{\beta} \right)  \mathcal I_0\left(\frac{2\sqrt{t\alpha}}{\beta} \right),\]
 \ie with mean $\alpha + \beta $ and variance $\beta^2 + 2\alpha \beta$. Then $X$ is sub-exponential with $\lVert X \rVert_{\psi_1} \leq 6\beta$.
\label{th:MR_subExp}
\end{proposition}

Therefore, sums of MR random variables enjoy the following concentration phenomenon shared by all sub-exponential random variables.
\begin{theorem}[\cite{vershynin2010introduction}]
Let $X_1,\ldots,X_m$ be independent $m$ centered sub-exponential random variables, and let $K = \max_i \lVert X_i\rVert_{\psi_1}$. Then, for every $\epsilon \geq 0$, we have
\begin{equation}
\textstyle P \left(  \sum_{i=1}^m X_i  \geq\epsilon m  \right) \leq  \exp [ -c \min(\frac{\epsilon^2}{K^2},\frac{\epsilon}{K})m],
\label{eq:subexp_decay}
\end{equation}
where $c>0$.
\label{th:subexp_decay}
\end{theorem}
We can thus bound the complementary CDF (\ie $1-CDF$) of $\sum_i X_i$ and state that for small values of the level $\epsilon$, sub-exponential random variables behave as Gaussian random variables while for larger values of $\epsilon$, they exhibit an exponential decay when $\epsilon$ increases. As detailed in Section~\ref{sec:sub_exp_processed_frame}, this change of behavior is critical to detect outliers to the speckle distribution, \eg planetary signals.

\subsection{Non-asymptotic analysis of the residual noise}
\label{sec:sub_exp_processed_frame}
 
The takeaway message of Proposition~\ref{th:MR_subExp} is that we can apply Theorem~\ref{th:subexp_decay} to characterize the tail decay of pixel intensities in the processed frame under the null-hypothesis (in the absence of a planetary signals). We first show how this can be done for a processed frame computed with the c-ADI algorithm. Then we argue that, even if in the case of PCA this analysis is more complicated and beyond the scope of this paper, we can still expect our analysis to hold.

For the c-ADI algorithm, $\bs L$ consists of $T$ copies of an image whose pixel intensities are the temporal median (or mean) of the data sequence of the corresponding trajectories. Hence the mean of the distribution is shifted towards zero but otherwise remains Modified Riccian and under the i.i.d. hypothesis, the sum along trajectories (step 3 in Figure~\ref{fig:ADI}) satisfies Equation~\eqref{eq:subexp_decay}. The processed frame is $f_g = \sum_{i}  s^{[g]}_i/T$ and for a trajectory $g$, under the null-hypothesis, we have
\begin{equation}
\textstyle P\left (  \frac{1}{T}\sum_{i}  s^{[g]}_i  \geq \epsilon  \right) \leq  \exp \left( -c \min( \epsilon^2/K^2, \epsilon/K)T \right).
\label{eq:behaviourReducedFrameK}
\end{equation}

From Equation~\eqref{MR_pdf} and Proposition~\ref{th:MR_subExp}, we note that if $I \sim \text{MR}(I_c,I_s)$, then $K \leq 6 I_s$ and the variance is equal to $\sigma_I^2 = I_s^2 + 2I_cI_s\geq K^2/36$. Therefore, $K \leq \sigma_I/6$ and the bound~\eqref{eq:behaviourReducedFrameK} implies
\begin{equation}
\textstyle P\left ( \frac{1}{T}  \sum_{i}  s^{[g]}_i   \geq \epsilon  \right) \leq  \exp \left( -\tilde c \min( \bar c  \epsilon^2/ \sigma_I^2, \epsilon/ \sigma_I)T \right),
\label{eq:behaviourReducedFrame}
\end{equation}
for some $\tilde c>0$ and $\bar c>0$. For small values of $\epsilon/ \sigma_I$, we do observe a Gaussian bound in Equation~\eqref{eq:behaviourReducedFrame}. On the contrary, for large values of $\epsilon/ \sigma_I$ the bound displays an exponential decay, \ie the tail decreases exponentially as $\epsilon$ increases. As the confidence level of the detection depends on the probability of having an outlier, it is important to take into account this phenomenon to avoid high false positive rates. Therefore we can reject the null-hypothesis by a careful selection of a threshold driven by the bound~\eqref{eq:behaviourReducedFrame}. This is the theoretical motivation of our detection map presented in Section~\ref{sec:STIMdetectionmap}.

\begin{figure}
  \centering
 { \includegraphics[scale=.8]{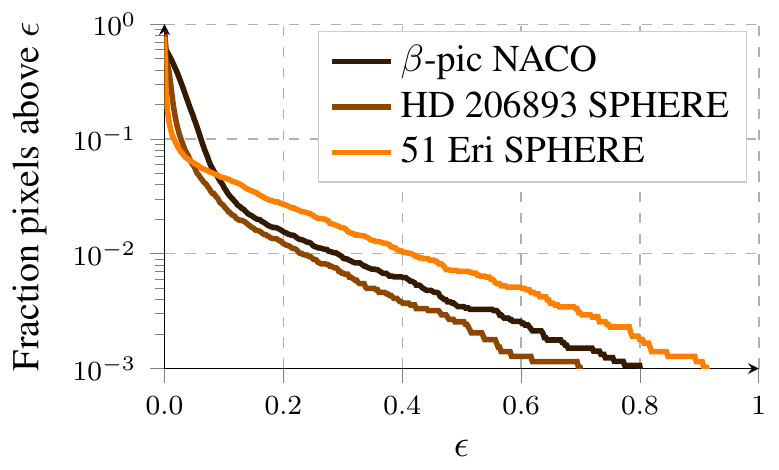}}
  \caption{The fraction of pixels in the opposite angles processed frame above a certain threshold $\epsilon$ as $\epsilon$ increases for three different datasets. We can see the fast decay for small values of $\epsilon$ followed by a smaller decay. Each opposite angles processed frame is obtained using one principal component.} \label{fig:sub_exp_vs_T}
\end{figure}

Moreover, the bound in Equation~\eqref{eq:behaviourReducedFrame} provides the information that the tail of distribution of the residual noise on the processed frame decays exponentially with the number of frames. Indeed, we see that the more i.i.d. random variables are added together, the larger the probability that the sum does not deviate from the mean. This effect is called the concentration of measure. The non-asymptotic nature of our analysis lies in this explicit dependence on the number of frames. 

To the best of the authors knowledge, these effects had not been fully theoretically assessed. This is the key theoretical contribution of the present paper as it explains the lower confidence level observed in the literature and it is the basis of the detection map proposed in Section~\ref{sec:STIMdetectionmap}.

\medskip
In the case of PCA, the theoretical analysis is more difficult and is left for future work. Nevertheless, we observed from the empirical Q-Q plots in Figure~\ref{fig:qq_plot} that PCA-generated processed frames exhibit residual noise with exponential tail decay. Furthermore, we argue that since $\bs L$ is built with few principal components is the low-rank structure of the data sequence, \ie that it captures the slowly varying parts of the data volumes, $\bs L$ remains close to the temporal mean. Hence the residual noise of a PCA-processed frame is expected to be similar to that of ADI-processed frame. We also note that PCA subtraction removes highly temporally correlated speckles, leaving lesser correlated speckles~\citep{mawet2014fundamental} and thus the i.i.d. hypothesis of Theorem~\ref{th:MR_subExp} is more likely to be verified for PCA. Indeed, we show in the Section~\ref{sec:noniid} that the non independence of the residual noise along trajectories results in a slow down of the concentration towards the mean and that the temporal correlation drops quickly as the number of principal components increases. 

To furthermore illustrate that we can use Equation~\eqref{eq:behaviourReducedFrame} to bound the probability that residual corruption on a PCA-processed frame reaches a value $\epsilon$, we  exhibit the sub-exponential behavior expected from Equation~(\ref{eq:behaviourReducedFrame}) in three different datasets using PCA in the reduction. To do so, we compute the empirical complementary CDF of the opposite angles processed frame of the three datasets. 
We count $n_\epsilon$ the number of pixels on the processed frame that are larger than $\epsilon$, for an $\epsilon$ ranging from zero to the maximal values taken on the processed frame. We display on Figure~\ref{fig:sub_exp_vs_T} the evolution of $n_\epsilon/n$ as $\epsilon$ increases for the three considered datasets. We can see that for all datasets the decay of the complementary CDF exhibits two distinct behaviors as expected from the bound of Equation~\eqref{eq:behaviourReducedFrame}. First a fast decay that is compatible with a quadratic (thus Gaussian) decay. Then a slower decay that displays a linear (thus exponential) trend.

For these reasons, we used a PCA to process the data throughout this paper and we characterised the resulting processed frame using our analysis. As demonstrated in Section~\ref{sec:simul}, we obtained convincing results regarding the capability of our method to detect exoplanets.

\subsection{Time-dependent residual speckles}
\label{sec:noniid}

As it is common in the literature, we assumed so far that the whitening effect of the reduction procedure is sufficiently strong to consider that the residual noise along trajectories is made of independent temporal components. That is the underlying assumption when one uses the CLT to state that the noise on the processed frame asymptotically follows a Gaussian distribution if the number of frames is large. In our analysis, we also sum random variables along trajectories and the bound given by Equation~\eqref{eq:behaviourReducedFrame} only holds for i.i.d. random variables $s^{[g]}_i$, \ie $s^{[g]}_i$ and $s^{[g]}_j$ are independent for $i \neq j$.

It is thus important to estimate the length of the temporal dependence of the residual speckle and its impact on the confidence level for the detection procedure. Due to its asymptotic nature, it is unclear how one can estimate the impact that the non-independence of the random variables has on the CLT, whereas this estimation is possible for our analysis. We here show how, under mild assumptions, the tail bound in Equation~\eqref{eq:behaviourReducedFrame} can be modified to account for the non-independence of the random variables $s^{[g]}_i$. We did not consider spatial dependence because it does not interfere with the assumptions of Theorem~\ref{th:subexp_decay}.

We are here interested in the typical temporal dependence along trajectories after subtraction of the model PSF \emph{in terms of number of frames} as it determines the number of i.i.d. random variables appearing in the sum of Equation~\eqref{eq:behaviourReducedFrame}. We assume a length of dependence $\tau$, such that $\tau$ successive elements of a trajectory are dependent. In other words, two elements of a trajectory $x_i$ and $x_j$ are independent if $\lvert i-j \rvert >\tau$ ($\tau$ being an integer). The result concerning the sub exponential decay of the tail is summarized in the following corollary, the proof is postponed to Appendix~\ref{app-proof-cor-1}.
\begin{corollary}
With a length of dependence $\tau$, Equation~\eqref{eq:behaviourReducedFrame} becomes
\begin{align}
 \textstyle &P\left (  \frac{1}{T}\sum_{i}  s^{[g]}_i   \geq \epsilon  \right) \leq \tau \exp \left( -\tilde c \min \left( \frac{\bar c\epsilon^2}{ \sigma_I^2}, \frac{\epsilon}{\sigma_I}\right)\frac{T}{\tau} \right).
\label{eq:corollary_slow_down}
\end{align}
\label{th:corollary_slow_down}
\end{corollary}

Corollary~\ref{th:corollary_slow_down} tells us that the temporal dependence slows down the concentration of measure around the mean.

It is in practice impossible to know the temporal dependence. Hence, we use the temporal autocorrelation (for each trajectory) after the subtraction of the model PSF in order to estimate the dependence of the residual noise. For a signal with non-correlated samples, the autocorrelation displays a large central peak surrounded by white noise. We averaged the normalised autocorrelation of all trajectories into a total autocorrelation. We define the correlation length as the number of frames for which the averaged autocorrelation is above a certain threshold.

\begin{figure}
  \centering
 {\includegraphics[scale=.8]{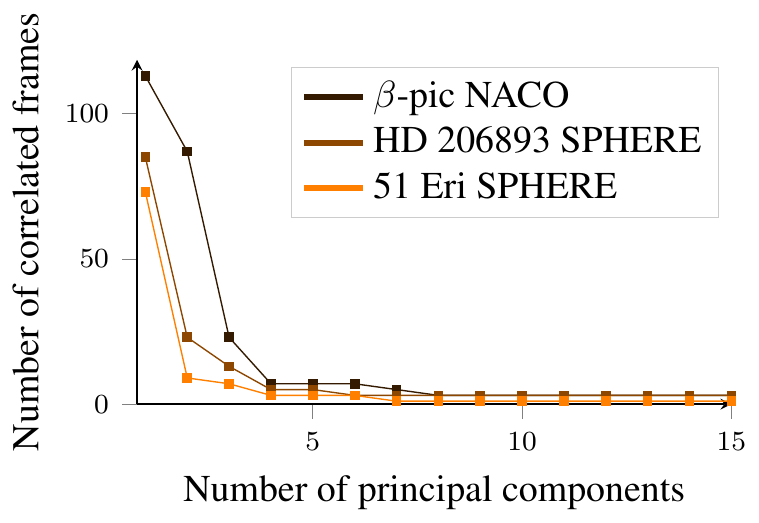}} 
\caption{Correlation length (in number of frames) with respect of the number of principal components used in the processing for different dataset. }
 \label{fig:mean_auto_corr_VS_rank} 
\end{figure}

We display on Figure~\ref{fig:mean_auto_corr_VS_rank} the evolution of the correlation length, using a threshold of $0.2$, for our three different datasets processed using PCA when the number of principal components increases. Although the typical correlation length decreases quickly as the number of principal components increases, its impact cannot be disregarded and the bound of Equation~\eqref{eq:corollary_slow_down} is more accurate than the bound of Equation~\eqref{eq:behaviourReducedFrame}. However, the estimation of the factor $\tau$ depends on the threshold one uses to compute the correlation length. Hence, the presence of $\tau$ (along with the other constants) in the bound of Equation~\eqref{eq:behaviourReducedFrame} hinders our capability to exactly estimate the inverse CDF of the distribution of the residual corruption. We show in Section~\ref{sec:simul} that the inverse CDF can be estimated from the opposite angles processed frame.

\section{Detection map}
\label{sec:stateart-detectionmap}
In this fourth section, we first briefly summarize a popular approach to compute a detection map from the processed frame, \ie the SNR$_t$ map \citep{mawet2014fundamental}, and list the shortcomings of this method. We then propose our novel detection map, the STIM map, and provide statistical guarantees for its subsequent thresholding based on our theoretical analysis.

\subsection{Current detection procedure for exoplanets direct imaging}
\label{sec:stateoftheart_detectionmap}

As the statistical properties of the processed frame are known to be radius dependent, the null-hypothesis is tested for each radius separately. And because of the diffractive effect, resolution elements of diameter $\lambda/D$ are considered. As there are fewer such resolution elements closer to the star, the confidence level of the detection suffers from small sample statistics.

\begin{figure}
  \centering
 { \includegraphics[width=.49\textwidth]{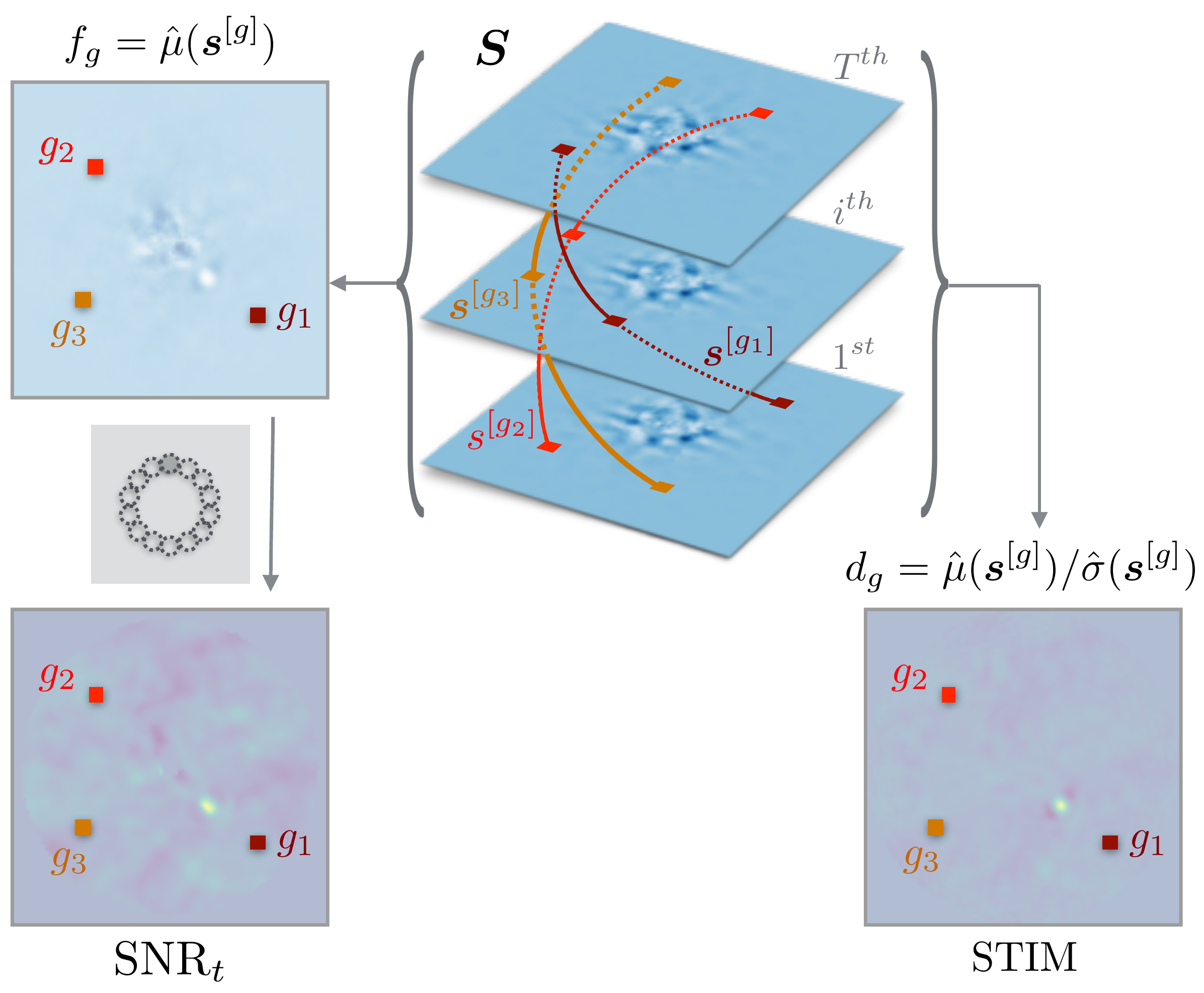}} 
  \caption{Illustrative views of the computations of both the SNR$_t$ and the STIM maps. We depicted three arbitrary trajectories $s^{[g_i]}$ and the corresponding elements $g_i$ of each frame to illustrate the notations used in the paper.
  The SNR$_t$ is built from the processed frame, comparing the statics of resolution elements located at the same radial distance from the center of the frame. The STIM map is constructed directly from the residual cube $\bs S$.} 
  \label{Illustration_SNR_vs_STIM}
\end{figure}

To deal with the small sample statistics and the high false positive rate obtained when relying directly on Gaussian confidence level, \cite{mawet2014fundamental} proposed another detection map based on the t-student statistics. The t-test is argued to be more robust with respect to small sample statistics and with respect to the deviation from a Gaussian distribution for the null-hypothesis. The test is defined by computing the following map from the pixels intensity on the processed frame
\begin{equation}
\textstyle \text{SNR}_t(x) = \frac{\bar{x}_1 - \bar{x}_2}{s_2 \sqrt{1 + \frac{1}{n_2}}},
\label{Eq_SNR_t}
\end{equation}
where $\bar{x}_1$ is the mean flux of intensity inside the circle surrounding the regarded pixel centered on $x$, $n_2$ is the number of the other $\lambda/D$ resolution elements located at the same radial distance from the center of the frame, and $\bar{x}_2$ and $s_2$ are respectively the mean fluxes and sample variances of all such resolution elements. The computation of the SNR$_t$ map is illustrated on Figure~\ref{Illustration_SNR_vs_STIM}. The resulting SNR$_t$ map and the processed frame from which it is built can be seen on Figure~\ref{fig:reduced_frame_and_SNR} for the three tested datasets. These images were built using the Vortex Image Processing (VIP) python package~\citep{gonzalez2017vip}\footnote{The code is publicly available at \url{https://github.com/vortex-exoplanet/VIP}.}.

\newcommand{\scaleExAll}{0.7}

\begin{figure*}
\begin{tabular}{lll}
  \centering
   \vspace{0.3cm}

    {\includegraphics[scale=\scaleExAll]{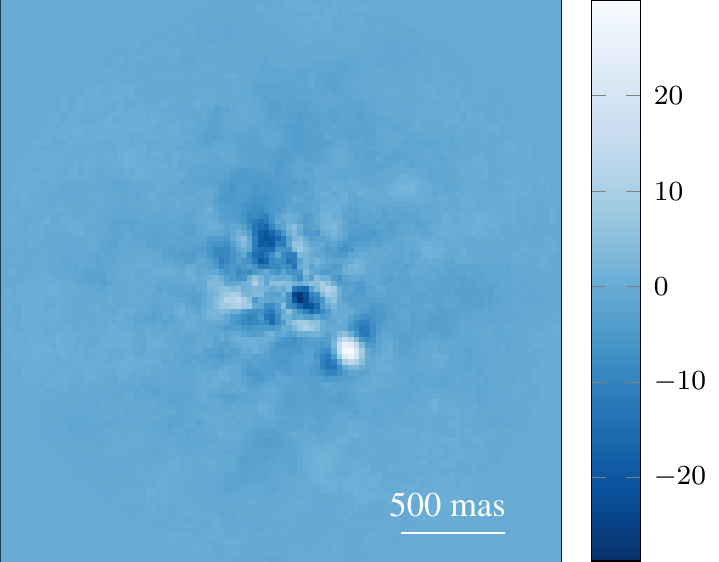}}&
    {\includegraphics[scale=\scaleExAll]{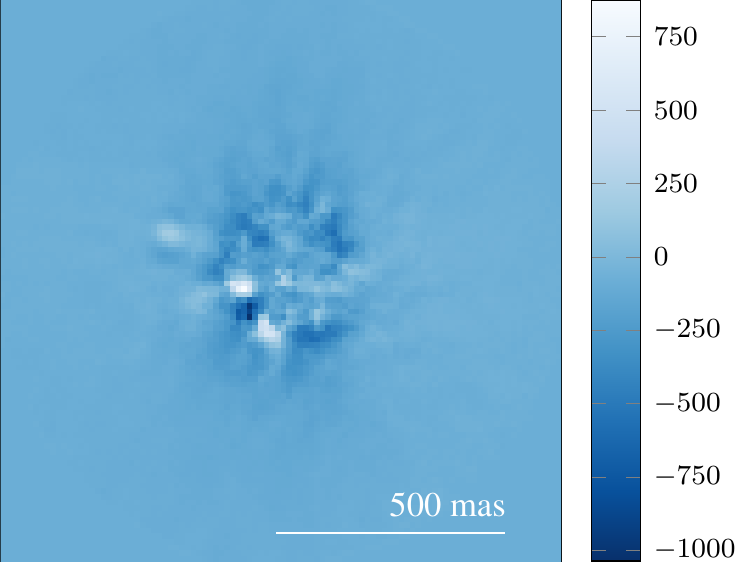}}&
    {\includegraphics[scale=\scaleExAll]{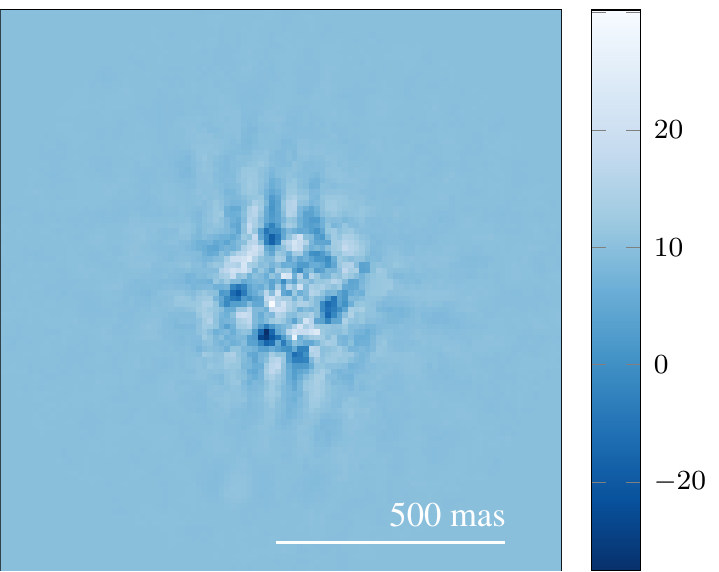}}\\
\vspace{0.3cm}
      {\includegraphics[scale=\scaleExAll]{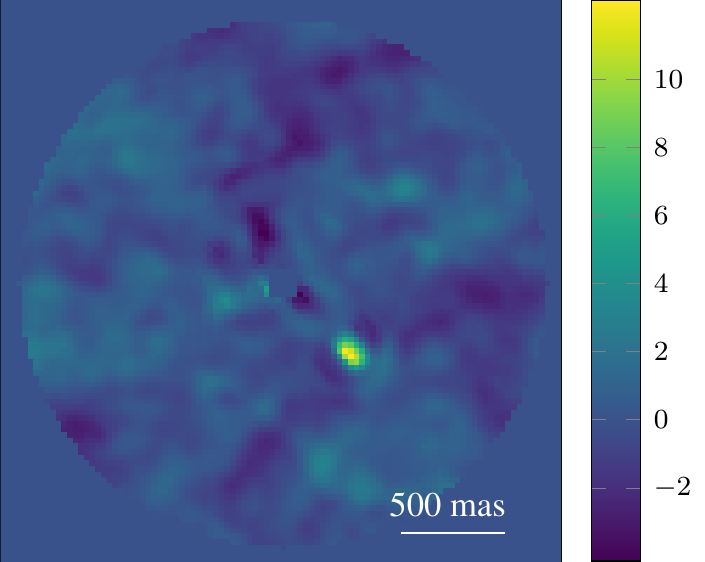}} &
          {\includegraphics[scale=\scaleExAll]{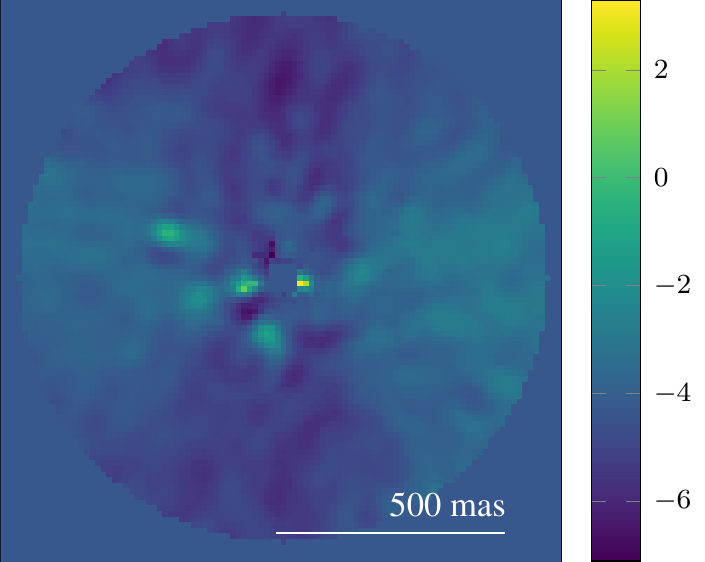}} &
        {\includegraphics[scale=\scaleExAll]{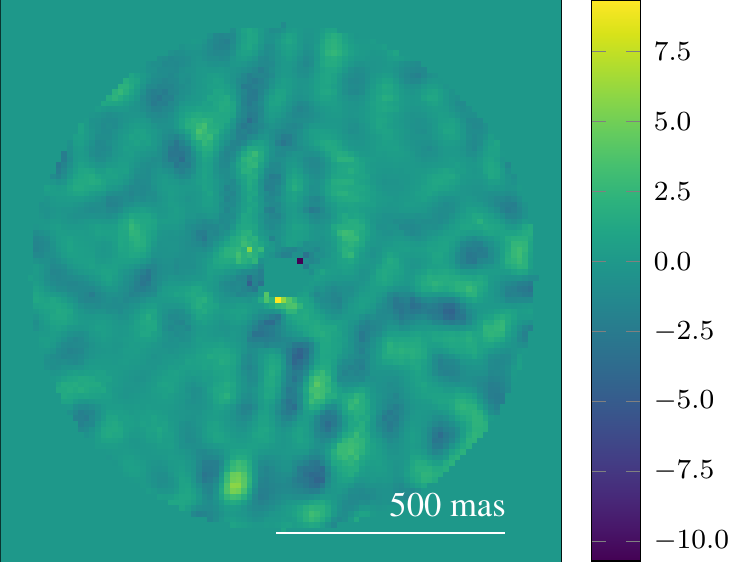}}\\ 
          
               {\includegraphics[scale=\scaleExAll]{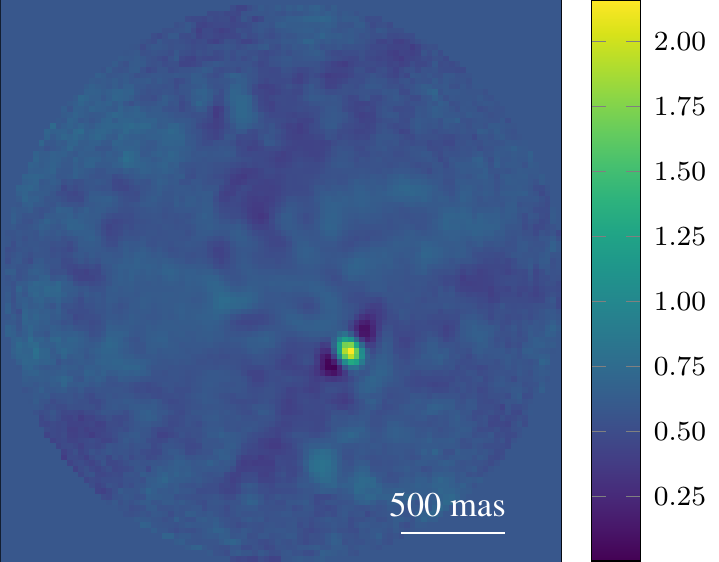}} &
      {\includegraphics[scale=\scaleExAll]{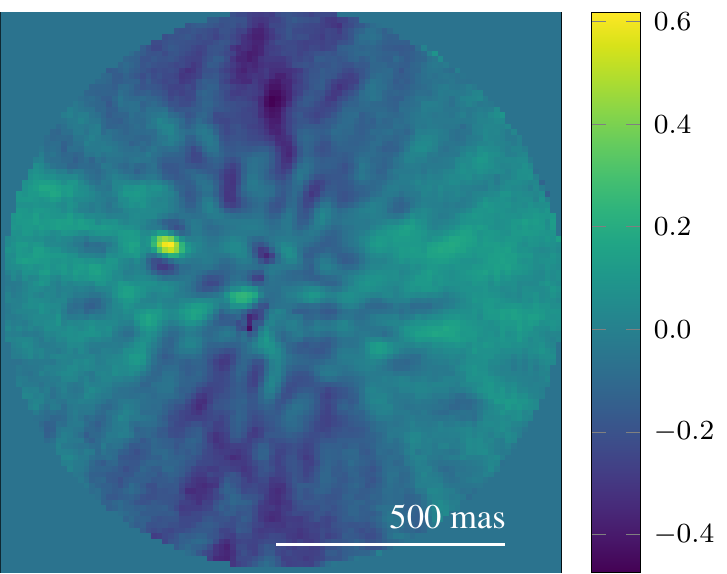}}&
      {\includegraphics[scale=\scaleExAll]{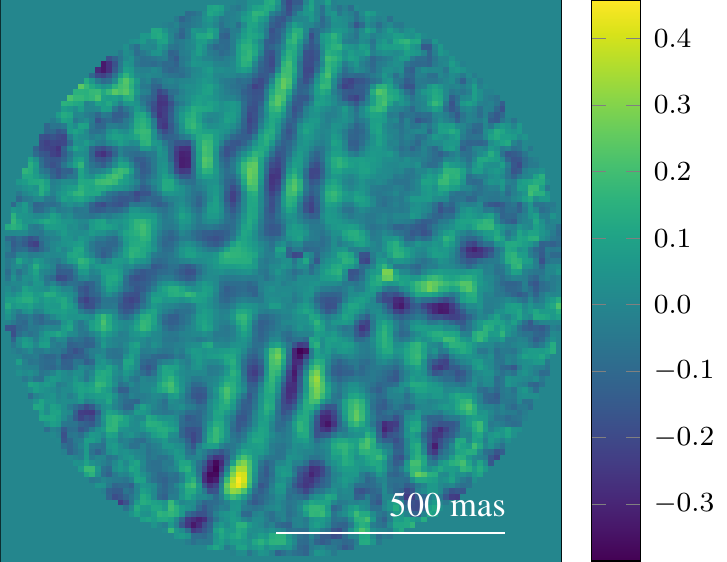}} 
 \end{tabular}
  \caption{Illustrative results for the three datasets: $\beta$-pic NACO (left), HD 206893 SPHERE (center) and Eri 51 SPHERE (right). Top: the processed frame, obtained with PCA, using respectively 3, 10 and 5 principal components. Middle: $\text{SNR}_t$ map. Bottom: STIM map.} 
 \label{fig:reduced_frame_and_SNR}
\end{figure*}

Despite its capacity to detect faint objects, \citep[see \eg][]{quanz2015confirmation,currie2015resolving}, there are two main drawbacks for this SNR$_t$. 

First, although the t-test is more robust with respect to deviation from a Gaussian distribution, this deviation is not quantified. We can hope that this distribution will be close to Gaussian if we add enough frames. But we have no knowledge of how robust the validity of the Gaussian hypothesis is with respect of the number of frames.

Second, the SNR$_t$ compares the statistical properties of a pixel against pixels at same angular separation from the star. If there exist multiple planets at the same radius, or extended structures such as circumstellar disks, the map will overestimate the noise at that radius, hence increasing the likelihood of a planet to be considered as a residual speckle. Furthermore, any test performed directly on the processed frame suffers from the small sample statistics when considering pixels near the center of the processed frame \ie when attempting to detect planets close to their host star.

Our objective is to define a robust detection procedure, still efficient at very close separation to the star in order to be sensitive to faint planetary signals all over the field of view, even at locations where the starlight residuals are very intense and varying fast (typically below a few $\lambda /D$ from the star). 

The proposed detection map is computed in the temporal domain and independently for each trajectory. Hence it has a sample statistics that does not depend on the radial distance from the center and the presence of multiple planets does not hinder the likelihood of a detection. Furthermore, the analysis of the distribution of the residual noise is non-asymptotic with respect to the number of co-added frames, consequently it is possible to estimate the confidence of a detection depending on the number of frames available.

\subsection{STIM: a time domain sub-exponential detection map}
\label{sec:STIMdetectionmap}

We now present our detection map and justify its use in the light of the results of the previous section. 
The presence of the standard deviation in the exponential in Equation~\eqref{eq:behaviourReducedFrame} induces a slower decay rate in areas of $\bs S$ with a larger standard deviation. Thus trajectories with larger temporal standard deviation are more likely to yield a significantly large value on the processed frame under the null hypothesis.
 For this reason, we propose to compute the map $\bs d$ whose components are given by 
\begin{equation}
 d_g = \frac{\hat{\mu}_g}{\hat{\sigma}_g},
\label{equation_detection_map}
\end{equation}
where 
$\hat \mu_g$ is the computed mean of trajectory $g$ in Equation~\eqref{processed_frame_mu} and $\hat \sigma_g$  its standard deviation; \eg the square root of its computed variance
\begin{equation}
 \hat \sigma_g^2 \equiv \hat \sigma(\bs s^{[g]})^2 = \frac{1}{T - 1} \sum_{i = 1}^{T} \left( s^{[g]}_i - \hat \mu_g\right)^2. 
 \label{eq_computed_sd}
 \end{equation}
We use $\bs d$ as a detection map and we name it the standardized trajectory intensity mean map or STIM map for short. Its computation is illustrated on Figure~\ref{Illustration_SNR_vs_STIM}. 

Another way to introduce the STIM map would be the following. The quantity $\hat \sigma_g$ is proportional to the standard deviation of the computed mean $\hat \mu_g$. Indeed, denoting the standard deviation of $\hat \mu_g$ by $\sigma(\hat \mu_g)$, it is given by 
\[ \sigma(\hat \mu_g) = \frac{\sigma(s^{[g]})}{\sqrt{T}}.\] 
Replacing the standard deviations by the estimator~\eqref{eq_computed_sd}, we get that the STIM map is given by
\[ d_g = \sqrt{T} \frac{\hat{\mu}_g}{\hat\sigma(\hat{\mu}_g)}. \]

\begin{figure}
  \centering
{\includegraphics[width=0.4\textwidth]{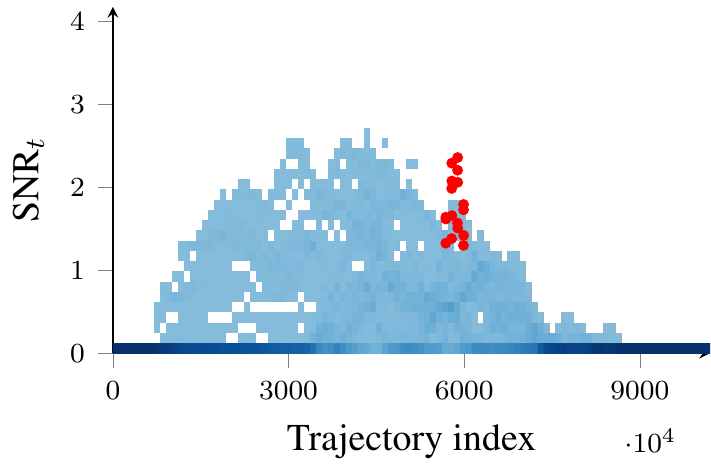}} \\
{\includegraphics[width=0.4\textwidth]{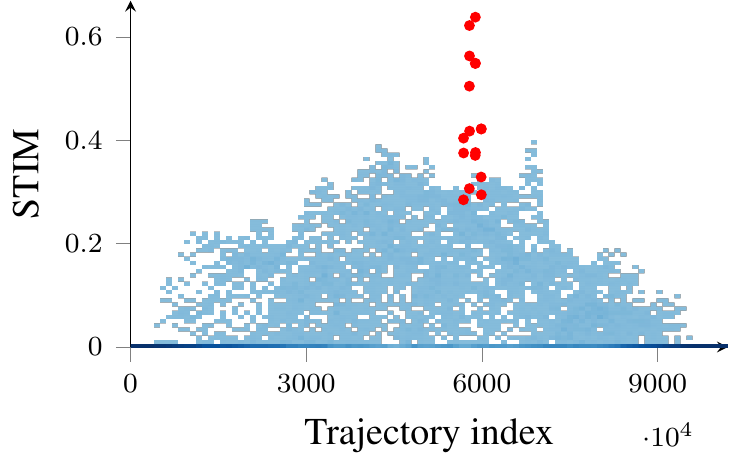}}    
   \caption{Comparison of the SNR$_t$ map (top) and the STIM map (bottom) in a one dimensional plot for the HD 206893 dataset. The planetary trajectories are represented with dots. Other trajectories are displayed as rectangles in a transparency fashion, the darker the more trajectories take this value. We can see that for the SNR$_t$ map, it is not possible to select a detection threshold such that planetary trajectories are the only to take values larger than the threshold. For the STIM map, such a threshold exists. In Section~\ref{sec:simul}, we describe how to set automatically this threshold from the complementary CDF.}
 \label{fig:example_plot_mu_sigma_map_sphere} 
\end{figure}

We display the STIM maps obtained for the three considered datasets on Figure~\ref{fig:reduced_frame_and_SNR}. On Figure~\ref{fig:example_plot_mu_sigma_map_sphere}, we also display a one dimensional plot of both the SNR$_t$ and the STIM map for the HD 206893 SPHERE dataset in order to better compare the behavior of the two maps. 

Intuitively, because residual speckles on $\bs S$ are spread on multiple trajectories, for a trajectory $g$ that is free of an exoplanet signal, the computed mean is expected to be small after the PCA subtraction. However, since most temporally correlated speckles are absorbed in $\bs L$ (see Section~\ref{PremNotations}), the computed variance of the intensities along the trajectory $g$ is expected to be large. Thus, we can expect to have $\hat \mu_g/\hat \sigma_g \ll 1$. 
On the other hand, since most the planets flux is still present on $\bs S$, for a trajectory $p$ containing a planet, we expect the computed mean to be large. Hence, we expect $\hat \mu_p/\hat \sigma_p$ to be larger than $ \hat \mu_g/\hat \sigma_g$.
 
We now show how we can use sub-exponentiality of the MR distribution to characterize the proposed detection map. Using the argument from Section~\ref{sec:MRsubexp}, we can apply Theorem~\ref{th:subexp_decay} with $ \epsilon \leftarrow \sigma_I\epsilon$ and Equation~\eqref{eq:behaviourReducedFrame} becomes
\begin{equation}
\textstyle P\left ( \frac{1}{T}  \sum_{i}  s^{[g]}_i   \geq \sigma_I \epsilon  \right) \leq  \exp \left( -\tilde c \min(\bar c \epsilon^2, \epsilon)T \right).
\end{equation}
As we noted, $\sigma_I$ depends on the separation from the star. We assume that there are enough frames so that  the computed standard deviation along a trajectory is a good estimate of $\sigma_I$ for that trajectory and we can set $ \epsilon \leftarrow \epsilon T \hat \sigma_g  \approx  \epsilon T \sigma_I$ and rearrange 
\begin{equation}
P\left ( \frac{  \sum_{i}  s^{[g]}_i }{T \hat \sigma_g}  \geq  \epsilon  \right) \leq  \exp \left( -\tilde c \min( \bar c \epsilon^2, \epsilon)T \right).
\end{equation}
By definition $\sum_i \bs s^{[g]}_i/T = \hat \mu_g$, hence
\begin{equation}
P\left (   \frac{\hat \mu_g}{ \hat \sigma_g}   \geq  \epsilon  \right) \leq  \exp \left( -\tilde c \min( \bar c \epsilon^2, \epsilon)T \right).
\label{proba_mu_sigma_large}
\end{equation}
In other words, the probability that the residual speckle noise reaches a large value on the STIM map decays exponentially with respect to $T$, the number of frames.

We argue that the proposed detection map does not suffer from the two drawbacks of SNR$_t$ described in Section~\ref{sec:stateoftheart_detectionmap}. 

First, by accessing the temporal domain, it is not impacted by the small sample statistics when considering planets close to their host (compared to the SNR$_t$ that only analyses the processed frame). Indeed, the sample statistics of the STIM map is the number of frames $T$ and is the same for all radii. 

We mention that if $T$ is small (or if the temporal correlation is strong, see Section~\ref{sec:noniid}),  $\hat \sigma$ is not necessarily a good estimator of the standard deviation. Nevertheless, the asymptotic analysis behind the state-of-the-art detection procedures suffers from the same small sample statistics. Indeed, if $T$ is small or strongly correlated, the hypothesis that the number of $i.i.d.$ random variables summed-up can be considered as infinite is severely hindered. Furthermore, the non-asymptotic nature of our analysis allows one to evaluate the quality of the estimator $\hat \sigma$ using standard statistical tools.

Second, by construction, the intensity of a given pixel on $\bs d$ does not depend on the intensity of other pixels. There is thus no influence of other planets that could be located on the same radius. We also note that the STIM map is easy to implement and fast to compute. For the illustrative examples of Figure~\ref{fig:reduced_frame_and_SNR}, the STIM maps only required a third of a second to be completed using a single CPU. A python implementation of the STIM map is available in the VIP toolbox.

The detection map suffers the same concentration of measure slow down as described in Section~\ref{sec:noniid}. With a length of dependence $\tau$, Equation~\eqref{proba_mu_sigma_large} becomes
\begin{equation}
 \textstyle P\left (  \frac{\hat \mu_g}{ \hat \sigma_g}   \geq  \epsilon  \right) \leq \tau \exp \left( - \tilde c \min( \bar c \epsilon^2, \epsilon)T/\tau \right).
 \label{proba_mu_sigma_large_slow_down}
\end{equation}

In the Section~\ref{set_detection_threshold}, we use the opposite angles detection map to estimate the typical values observed on the detection map under the null hypothesis. As with the opposite angles procedure the planetary signal is significantly reduced while preserving the speckle temporal dependence, the concentration of measure slow down of Equation~\eqref{proba_mu_sigma_large_slow_down} is then accounted for.

Before moving on, we note that the bound similar to the one  in Equation~\eqref{proba_mu_sigma_large}  would still hold if the mean is replaced by the median~\citep{wainwright2019high}. However, when performing experiments with median, we did not find significant improvement over the mean (See Appendix~\ref{app-compMedianMean}).

\section{Experiments and discussion}
\label{sec:simul}
In this section, we provide some numerical experiments to demonstrate the capabilities of our approach.

\subsection{Detection threshold estimation}
\label{set_detection_threshold}

We know from sub-exponentiality of the MR distribution that the ratio $\sfrac{\hat \mu_g}{\hat \sigma_g}$ is close to zero with high probability under the null-hypothesis. Unfortunately, the unknown constants $\tilde c$ and $\bar c$ in Equation~\eqref{proba_mu_sigma_large} and the potential temporal dependence in the residual noise prevents us to leverage this bound to determine a detection threshold rejecting $H_0$.
We here propose to select a detection threshold by estimating the complementary CDF of the residual noise on the processed frame, \ie estimating $P(\hat \mu_g / \hat \sigma_g > \epsilon | H_0)$. If this function is known, a threshold can be set so that it is unlikely that the residual noise reaches its value. As we do not have access to the true complementary CDF of the residual noise, we propose to estimate it from the empirical CDF computed from the opposite angles detection map. As outlined previously, the temporal dependence of the speckle noise is the same as in the initial dataset. Therefore, the concentration of the $\hat \mu/\hat \sigma$ around $0$ is also preserved with the same potential concentration of measure slow down (see Section~\ref{sec:noniid}). Mathematically, $P(\hat \mu_g / \hat \sigma_g > \epsilon | H_0) \approx P(\hat \mu_{\bar g} / \hat \sigma_{\bar g} > \epsilon )$ holds because $H_1$ does not arise in the opposite angles processed frame.
We estimate the complementary CDF as the number of pixels taking a value larger than a threshold for increasing values of this threshold. We thus compute $n_\epsilon = \sum_g \chi_+( \hat \mu_g/\hat \sigma_g - \epsilon)$, where $\chi_+(x) = 1$ if $x>0$ and $0$ otherwise. If n is sufficiently large, we have  $n_\epsilon/n \approx P(  \hat \mu_g/\hat \sigma_g > \epsilon)$ and $n_\epsilon/n$ is then a good approximation of the complementary CDF.

\begin{figure}
  \centering
{ \includegraphics{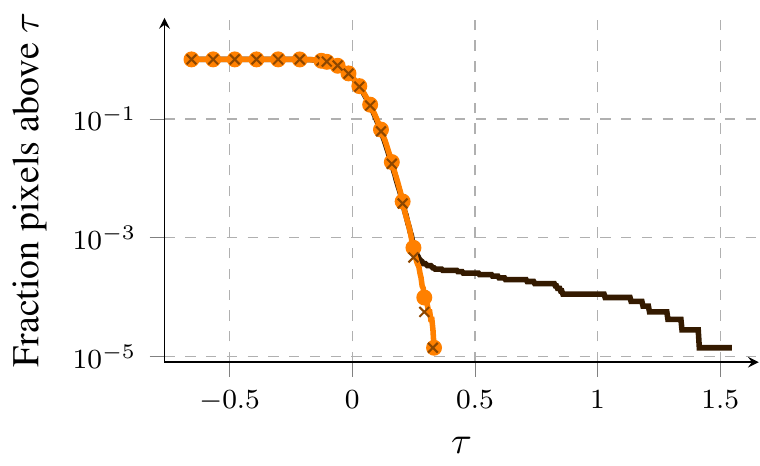}}
  \caption{Plot of the fraction of pixels in the STIM map above $\tau$ in three different settings: (i) dark, the planet is present, (ii) medium, the planet is removed with the negative companion injection method, and (iii) light, the planet is present but the opposite parallactic angles are used in the processing. As the decay is indistinguishable for the cases (ii) and (iii), we added dots (light) and crosses (medium) to ease the comparison. On the other hand, the decay is significantly different when a planet is present. These results are obtained on the $\beta$-pic dataset processed using a PCA with 10 principal components.}
  \label{Inverse_CDF_comparison}
\end{figure}

To show that we can indeed use the opposite parallactic angles technique to estimate the residual noise level, we compare the empirical complementary CDF of the detection map in three different cases: (i) the planet is present in the dataset, (ii) the planet signal is strongly attenuated using the negative companion injection method, and (iii) the planet is present but we used the opposite parallactic angles in the processing. We have observed that the curves in cases (ii) and (iii) are practically indistinguishable, hence the complementary CDF of the residual noise can be estimated from the detection map obtained with the opposite parallactic angles. We display the comparison for the $\beta$-pic dataset on Figure~\ref{Inverse_CDF_comparison} where we see that $\tau=0.5$ is an adequate detection threshold. On Figure~\ref{fig:fake_comps}, we see that this threshold allows to detect the three injected companions without any false positive.

\begin{figure}
  \centering
  { \includegraphics[width=.4\textwidth]{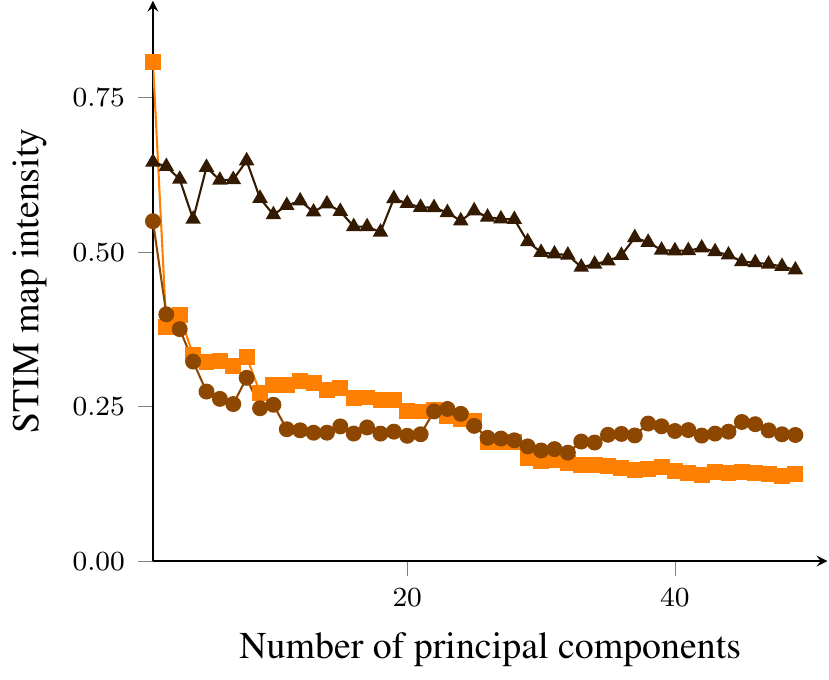}}\\
{ \includegraphics[width=.4\textwidth]{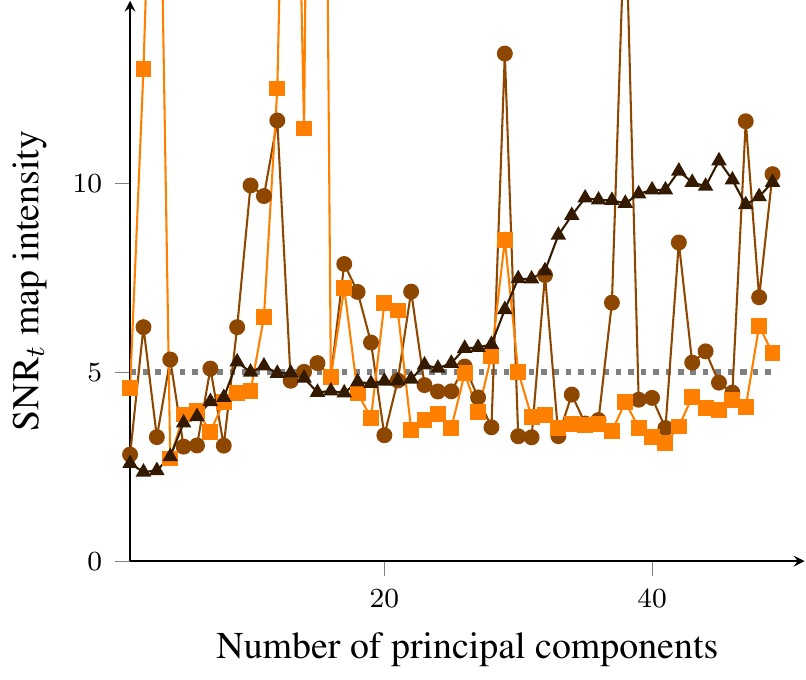}}
  \caption{Detectability level for STIM map (top) and for SNR$_t$ map (bottom) obtained with the HD 206893 dataset. In dark (triangle), intensity at planet location, and in light (rectangle), detectability level estimated by inverse trajectories. In medium (dots), detectability level estimated by looking at the second largest spot on the detection map. On the SNR$_t$ map we overplotted in gray dashed the $5\sigma$ threshold. The STIM map detects the planet for all rank larger than 1 while SNR$_t$ map requires large number of principal components to yield sufficiently large values of the SNR.} \label{fig:detectability_vs_rank}
\end{figure}

We also show the sensitivity of the threshold with respect of the number of principal components used in the PCA. We compare the intensity of the planetary signal on the detection maps with \textit{(i)}~the largest intensity observed on the opposite angles detection map and \textit{(ii)}~the largest intensity on the the detection map under the null-hypothesis. Figure~\ref{fig:detectability_vs_rank} displays the evolution of these three quantities as the number of principal components increase from 1 to 50, for the STIM and SNR$_t$ maps obtained with the HD 206893 dataset. For the STIM map, the opposite angles detection map yields a good approximation of the maximal noise intensity on the actual detection map. Furthermore, this approximation can be used to estimate a detection threshold such that the planetary signal is detected without false positives. In contrast, for the SNR$_t$, the maximal values on the opposite angles map is not a good estimate for the maximal value of the detection map under the null-hypothesis. In addition, for a wide range of numbers of principal components, the SNR$_t$ map is unable to detect the planet without false positive. We observed a similar trend for the other datasets.

\subsection{Receiver operating characteristic curves}

Receiver operating characteristic (ROC) curve is a widely used tool to compare classifiers. It consists of plotting the true positive rate (TPR) as a function of the false positive rate (FPR). A good classifier yields a good trade-off between large TPR for small FPR.

We propose to build a ROC curve that is localized following the procedure described in~\cite{gonzalez2017supervised}. The resulting ROC curve is close to an alternative free-response operating characteristic (AFROC) curve where one plots the fraction of objects detected versus the fraction of images with one or more false positives~\citep{metz2006ROC}. Given a detection threshold, we say there is a TP if there is, within a $\lambda/D$ diameter area around the planet's position, one pixel whose intensity is above the threshold. We count the number of FP's as the number of $\lambda/D$ circular areas that contain at least one pixel above the threshold, the experiment is repeated 100 times. We computed the TPR and the FPR for a given radius and for a given intensity range. 

\begin{figure}
  \centering
  { \includegraphics[width=0.4\textwidth]{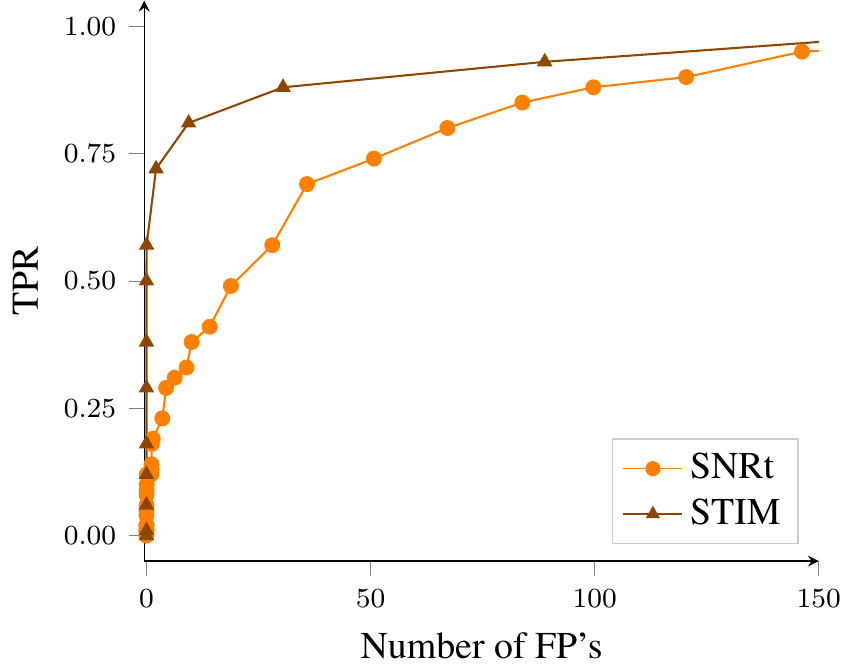}}\\
  { \includegraphics[width=0.4\textwidth]{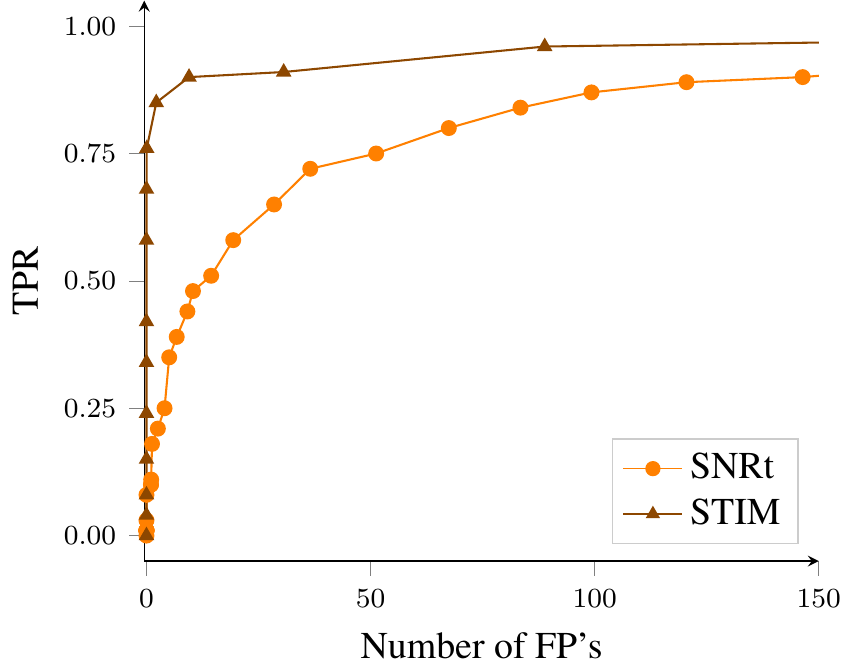}}
  \caption{ROC curves obtained for 100 fake companion injections, all processed using 4 principal components. Top: at radial separation $250 \text{ mas}$ with contrast between  $2.75\times 10^{-5}$ and $4.12\times 10^{-5}$. Bottom: at radial separation $375 \text{ mas}$ with contrast between $1.37\times 10^{-5}$ and $2.75\times 10^{-5}$.} \label{fig:ROC_combined_all_IR_rank.png}
\end{figure}

We display in Figure~\ref{fig:ROC_combined_all_IR_rank.png} the ROC curves obtained for the $\beta$-pic dataset with synthetic planets injected at small inner working angles (top) and high contrast (bottom). In both cases, we observe a gain in terms of true positive rates vs false positive rate ratio. For the small inner working angles case, the true positive rate for no false positive is $9\%$ for SNR$_{t}$ map, whereas the STIM map reaches $57\%$ true positive rate for no false positive. For high contrast, these percentages are respectively $10\%$ and $76\%$. We conclude that in both cases, a significant improvement is achieved.

\subsection{Example with multiple planets at same radius }

In Section~\ref{sec:stateart-detectionmap}, we stated that when multiple planets are located at the same radial distance from the host star, or when extended structures are present, the SNR$_t$ tends to overestimate the residual noise and this hinders its capability to detect planets. We illustrate this effect in this first experiment. We inject three planets at the same radial separation in a planet-free dataset and show how the SNR$_t$ and STIM maps behave. 

We perform this experiment on the $\beta$-pic dataset using the opposite angles to significantly reduce the effect of the known companion. Then we injected three synthetic planetary signals, at a radial separation of 352 mas and position angles of $133\degree$, $270\degree$, and $15\degree$ with contrast of $4.41\times 10^{-5}$, $4.81 \times 10^{-5}$, and $3.73 \times 10^{-5}$ respectively \citep[which is below the $5\sigma$ detection limit presented in][]{absil2013searching}. We processed the data with a PCA, using 5 principal components. Figure~\ref{fig:fake_comps} shows the processed frame, the SNR$_t$ map, and the STIM map. We see that no planet is found above the noise level in the SNR$_t$ map whereas all three planets are above the noise level in the STIM map.

\newcommand{\scaleExampleFakeComps}{0.65}

\begin{figure*}
  \centering
\begin{tabular}{llll}
   \vspace{0.3cm}

    {\includegraphics[scale=\scaleExampleFakeComps]{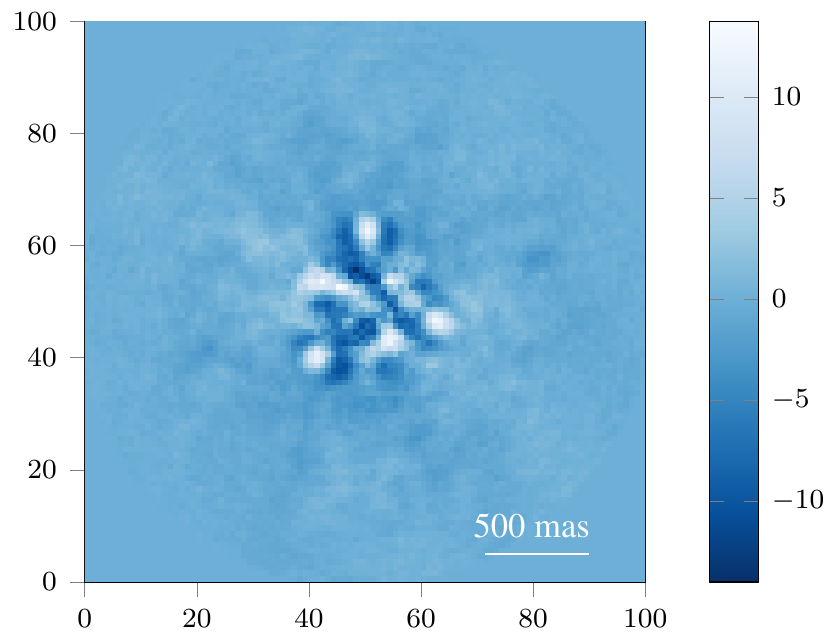}}&
    {\includegraphics[scale=\scaleExampleFakeComps]{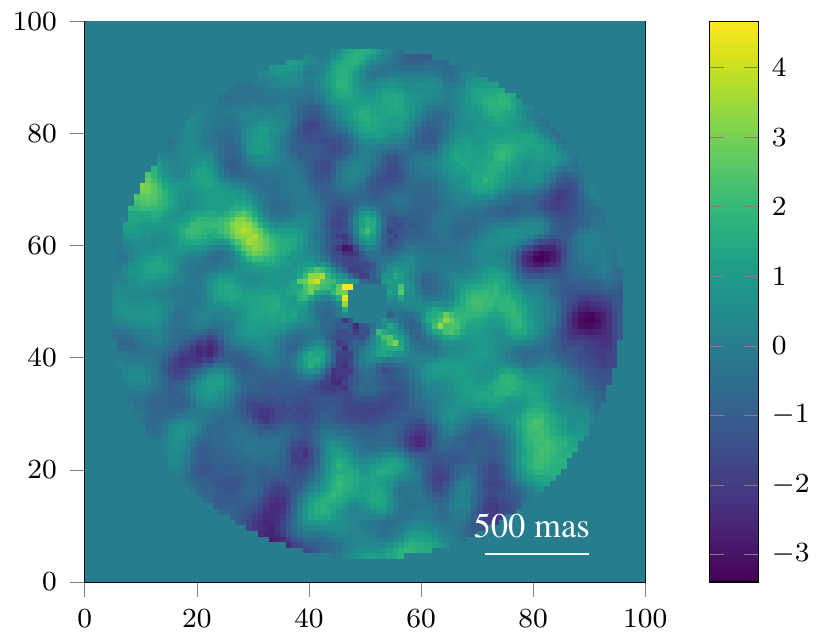}}&
    {\includegraphics[scale=\scaleExampleFakeComps]{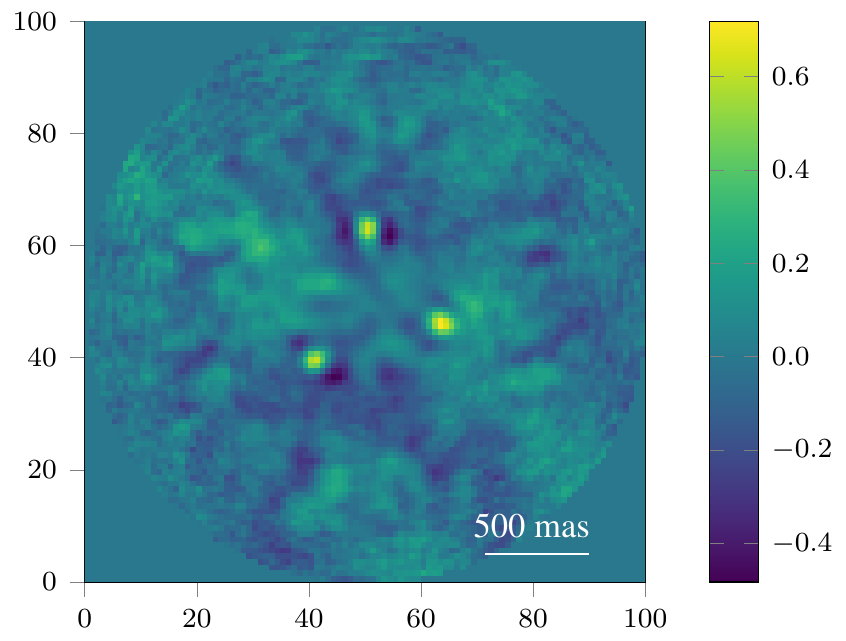}}\\
 \end{tabular}
      {\includegraphics[width=0.43\textwidth]{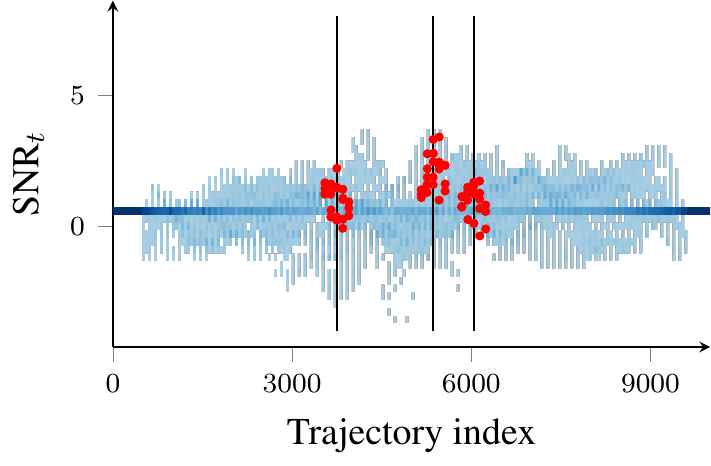}} 
      \hspace{0.5cm}
  {\includegraphics[width=0.43\textwidth]{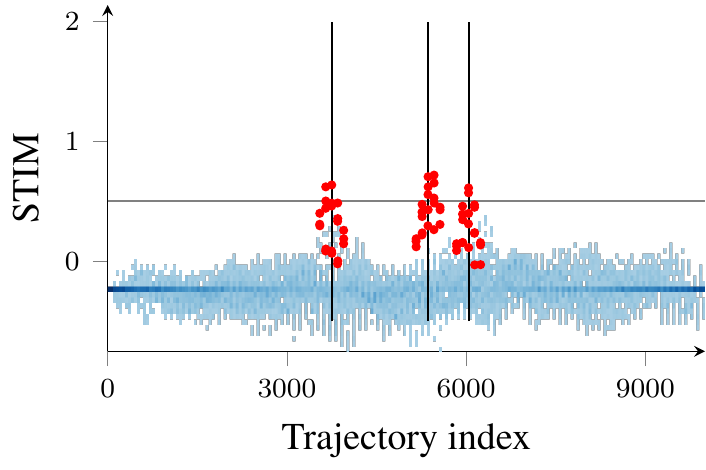}} 
  \caption{Results with 3 synthetic exoplanets injected at the same separation in the $\beta$-pic NACO data. Top row: processed frame using PCA with 5 principal components (left), $\text{SNR}_t$ map (center), and STIM map (right). Bottom row: intensities of the $\text{SNR}_t$ map (left) and of the STIM map (right) for each trajectory. For the $\text{SNR}_t$ map, we see that no planet stands above the noise level, while for the STIM map, all three planets are detected without false positive when using the $0.5$ threshold derived in Section~\ref{set_detection_threshold}.}  
  \label{fig:fake_comps}
\end{figure*}

\section{Conclusion and future work}
\label{sec:conclu}

In this paper, we first showed empirically that the tail decay of the residual noise on the processed frame is better explained by a Laplacian distribution than by a Gaussian distribution. From this qualitative observation, we then theoretically proved that the MR distribution is sub-exponential which means that its tail decays as an exponential. We used non-asymptotic statistical analysis to show that the tail distribution of the residual noise on the final post-processed frame indeed decays as an exponential. Compared to the commonly used CLT approach, this non-asymptotic analysis directly estimates the sensitivity of the detection procedure with respect to the number of frames we have in a given dataset and with respect to the temporal correlation of the residuals. 

Based on our statistical analysis, we introduced a novel detection map, called STIM map, and studied its theoretical properties. We then used thorough numerical experiments to demonstrate its capabilities on real data. We also provided a method to automatically estimate the detection threshold from any dataset. This detection map thus enables direct and automatic detection in PCA-processed frames with a higher true positive to false positive ratio than state-of-the-art SNR$_t$ map used for exoplanet imaging today. In particular, our performance analysis showed that the proposed STIM map reaches a significant gain in terms of detection, especially at small inner working angles where the poor field rotation and the high speckle noise variance makes it very difficult to extract signals and reveal the presence of fainter signals. This usual limitation is bypassed thanks to the time-domain approach of the STIM map that, to the best knowledge of the authors, has never been used in the framework of high contrast imaging post-processing.

On top of being directly used after speckle subtraction algorithms, our analysis has implications on other algorithms. For instance inverse problems approaches, such as ANDROMEDA or FMMF, rely on a maximum a posteriori (MAP) estimate thus the knowledge of the actual distribution of the residual noise allows to take it into account when solving the problem.
Indeed, these authors assume that the residual distribution is Gaussian, which implies the use of an $\ell_2$-norm estimation. We have shown in this paper that the decay of the tail is exponential and better accounted for by a Laplace distribution. It turns out that the optimal norm to use in the MAP estimation for the Laplace distribution is an $\ell_1$-norm. This solution is being implemented in such methods and published in a forthcoming paper. 

Future work would involve studying the specific case of extended features, such as faint debris disks with sharp edges \citep[see \eg ][for a gallery]{lee2016primer} or bright protoplanetary disks showing blunt structures such as spiral arms \citep[see \eg ][]{benisty2015spirals,benisty2017spirals}. 
Another step is to apply this method on images taken using different diversity such as the spectral diversity provided by integral field spectrograph, where the planet extraction and characterization is difficult due to the degeneracy of the obtained signal-to-noise ratio with planetary spectrum. 

\section*{Acknowledgements}

We would like to thank the anonymous reviewer and Tim Brandt for their insights and comments that greatly improved the manuscript.

We thank our colleagues, including V. Cambareri, K. Degraux, A.-L. Maire P.-A. Absil, and X. Lambein for the many fruitful discussions that greatly contributed to the research. 

We also would like to thank Thomas Henning for hosting Beno\^it Pairet at the Max Planck Institute for Astronomy (MPIA) to make progress on this work and use data sets from various instruments.


\newpage
\bibliographystyle{mnras}
\bibliography{biblio}

\newpage
\appendix

\begin{table*}
\centering
\caption{Description of the three datasets used in this paper. The total number of images constituting the data cube is noted $N_{\text{images}}$. The integration time used for each exposure of this data cube is noted $t_{\text{int}}$. The average turbulence coherence time during the observation is noted $t_{\text{coh}}$.}
	\label{table:list_data_set}
	\begin{tabular}{l c c c c c c c c} 
		\hline
		Name & Date & Instrument & Coronagraph & Filter & $N_{\text{images}}$ & Total field rotation & $t_{\text{int}}$ & $t_{\text{coh}}$ \\
             &      & (VLT)      &             &   		&			   & [deg] 			 & [s] 		  & [ms] \\
		\hline
		$\beta$-pic & Fev 2013 & NACO 		  & AGPM & Lp ($3.5$-$4.1\;\mu$m)  & 612 & 83.0 & 8 & 2.1 \\
        HD 206893 	& Oct 2015 & SPHERE-IRDIS & APLC & H ($1.48$-$1.77\;\mu$m) & 570 & 49.3 & 4 & 2.8 \\
        51 Eri	 	& Sep 2015 & SPHERE-IRDIS & APLC & K1 ($2.0$-$2.2\;\mu$m)  & 256 & 41.6 & 16 & 6.2 \\
        \hline
	\end{tabular}
\end{table*}

\section{Presentation of the dataset used in this paper}
\label{app-data_set}
Throughout this paper, we used three different representative dataset to test our approach: (1) Beta Pictoris (HIP 27321) observations using the VLT/NaCo instrument, published in \cite{absil2013searching} (ESO program ID 60.A-9800), (2) HD 206893 (HIP 107412) observations using the VLT/SPHERE-IRDIS instrument, published in \cite{milli2017discovery} (ESO program ID 96.C-0388) and (3) 51 Eridani (HIP 21547) observations using the VLT/SPHERE-IRDIS instrument, published in \cite{samland2017spectral} (ESO program ID 095.C-0298). Each target has one companion discovered by imaging: (1) $\beta$-pic~b \citep{lagrange2010giant} is a 8 to 15 Jupiter mass exoplanet orbiting its host star at 5 to 10 au, (2) HD 206893~b \citep{milli2017discovery} is a 12 to 50 Jupiter mass companion orbiting its host star at 10 to 15 au \citep{delorme2017charac} and (3) 51 Eri~b \citep{macintosh2015discovery} is a 2 to 10 Jupiter mass companion orbiting its host star at 10 to 14 au \citep{samland2017spectral}.

Table~\ref{table:list_data_set} gathers the main information about each dataset taken with instrument located at the ESO Paranal observatory. The three dataset make use of a coronagraph: the $\beta$-pic data are taken with an Annular Groove Phase Mask (AGPM) vector vortex coronagraph \citep{mawet2005annular} optimized in L' band \citep{mawet2013agpmlband} and the two SPHERE datasets are taken using the Apodized Lyot Coronagraph \citep[APLC,][]{soummer2005aplc} optimized for the YJH bands (focal mask diameter of 185 mas) and including a Lyot stop \citep{boccaletti2008aplc}.

\section{Proof of sub-exponentiality of MR distribution}
\label{ProofSubExpMRProp}
We prove Proposition~\ref{th:MR_subExp}, \ie let $X \sim \text{MR}(\alpha,\beta)$ with 
\[ \text{MR}(\alpha,\beta) \sim \frac{1}{\beta} \exp \left(-\frac{t+\alpha}{\beta} \right) \mathcal I_0\left(\frac{2\sqrt{t\alpha}}{\beta} \right),\]
 then $X$ is sub-exponential with $\lVert X \rVert_{\psi_1} \lesssim 6\beta$.

\begin{proof}
The moments of $X$ are given by 
\[\textstyle \mathbb E X^p = \int_0^{+\infty} t^p\frac{1}{\beta} \exp \left( -\frac{t+\alpha}{\beta}  \right)\mathcal I_0\left(\frac{2\sqrt{t\alpha}}{\beta} \right) dt,\]
by change of variable $t=s^2$, $dt = 2 s ds$, we get 
\begin{align*}
&\textstyle \mathbb E X^p = \\
& \textstyle \int_0^{+\infty} s^{2p} \left( \frac{2s}{\beta}  \exp \left( -\frac{s^2+\alpha}{\beta}  \right) \mathcal I_0\left(\frac{2s\sqrt{\alpha}}{\beta} \right) \right) ds\\
&\textstyle = \frac{1}{\sqrt{2}} \textstyle \mathbb EY^{2p},
\end{align*}
 where $Y$ follows a Rice distribution~\citep{papoulis2002probability}: 
\[ Y \sim \text{Rice}(\nu,\sigma) \sim \frac{s}{\sigma^2} \exp \left( -\frac{s^2+\nu^2}{2\sigma^2}  \right)\mathcal I_0\left(\frac{s\nu}{\sigma^2} \right)\]
with $\nu = \sqrt{\alpha}$ and $\sigma = \sqrt{\frac{\beta}{2}}$.
We thus have to prove that $\frac{1}{p}( \textstyle \mathbb EY^{2p})^{1/p}$ is bounded.  The raw moment of $Y$ are given by\footnote{\url{https://reference.wolfram.com/language/ref/RiceDistribution.html}}
\[ \textstyle \mathbb EY^{2p} = \sigma^{2p} 2^{p}\Gamma(1+p)L_{p}(-\frac{\nu^2}{2\sigma^2}),\] where $\Gamma(\cdot)$ is the gamma function and $L_{\gamma}(\cdot)$ is the Laguerre polynomial of degree $\gamma$.

Since for $x>0$, $\Gamma(x)< \sqrt{2\pi} e^{-(11/12) x} x^{x-1/2} $ [Section~5.6]~\citep{NIST:DLMF}, we find, for $p\geq 1$,
\[ \Gamma(1+p)< \sqrt{2\pi}  (1+p)^{\frac{1}{2}+p}, \]
where we use the fact that $e^{-(11/12) x} < 1$.
Moreover, for $n\in \mathbb N$, $L_{n}(\cdot)$, [Equation~18.14.8]~\citep{NIST:DLMF}:
\[\lvert L_n(x) \rvert  \leq \frac{1}{n!} \exp(x/2)  ,\]

Gathering all these observations yields: 
\[ \textstyle \mathbb E(Y^{2p}) < \left(\sigma^{2p}  \frac{2^{p}}{p!} \sqrt{2\pi}  (1+p)^{\frac{1}{2}+p}  \exp(-\frac{\nu^2}{4\sigma^2})\right).\]
Thus for $p\geq 1$,
\[ \textstyle 
p^{-1} (\mathbb EY^{2p})^{\frac{1}{p}}  < p^{-1} \left(\sigma^{2p}\frac{2^{p}}{p!}  \sqrt{2\pi}  (1+ p)^{\frac{1}{2}+p} \right)^{\frac{1}{p}}.\]
Recalling that $\mathbb EY^{2p} = \sqrt{2} \mathbb \mathbb EX^{p}$, we get 
\begin{align*}
p^{-1} (\mathbb EX^{p})^{\frac{1}{p}}  &< p^{-1} \left(\sigma^{2p} \frac{2^{p}}{p!}  \sqrt{\pi}  (1+ p)^{\frac{1}{2}+p} \right)^{\frac{1}{p}}\\
& < p^{-1} \left(\sigma^{2p} \frac{2^{p}}{p!}  \sqrt{\pi}  (1+ p)^{\frac{1}{2}} (1+ p)^p \right)^{\frac{1}{p}}\\
& <  \sigma^{2} 2  \sqrt{\pi}^{\frac{1}{p}}  \left(\frac{\sqrt{1+ p}}{p!}\right)^{\frac{1}{p}} (\sfrac{1}{p}+ 1) .
\end{align*}
To find a bound for $\left(\frac{\sqrt{1+ p}}{p!}\right)^{\frac{1}{p}} $, we consider two cases. First, the case $\frac{\sqrt{1+p}}{p!}\leq 1$, then $\left(\frac{\sqrt{1+ p}}{p!}\right)^{\frac{1}{p}} \leq1$. Otherwise, $\left(\frac{\sqrt{1+ p}}{p!}\right)^{\frac{1}{p}} \leq \frac{\sqrt{1+ p}}{p!} \leq   \frac{\sqrt{1+ p}}{p} = \sqrt{1/p^2+ 1/p}$. Then we observe that $\sqrt{1/p^2+ 1/p}$ is a monotonically decreasing function reaching its maximum for $p=1$ with value $\sqrt{2}$. Hence in both cases, we conclude $\left(\frac{\sqrt{1+ p}}{p!}\right)^{\frac{1}{p}} \leq \sqrt{2}$.
For $ (\sfrac{1}{p}+ 1)$, we also use the fact that it is a monotonically decreasing function reaching its maximum for $p=1$ with value 2. 
Collecting these results and observing that $\sqrt{\pi}^{1/p} \leq \sqrt{\pi}$, we get
\[ \lVert X \rVert_{\psi_1}  \leq \sigma^2 4 \sqrt{\pi} \sqrt{2}  = \beta 2 \sqrt{2 \pi} < 5.02 \beta < 6 \beta. \]
Hence we have shown that $X \sim \text{MR}(\alpha,\beta)$ is sub-exponential with $\lVert X \rVert_{\psi_1}< 6 \beta$.

\end{proof}

\section{Proof of the concentration of measure slow down induced by non independence}
\label{app-proof-cor-1}
We prove Corollary~\ref{th:corollary_slow_down}, we first present a proposition.
\begin{proposition}
Assume $B = \sum_{i=1}^\Delta A_i$ and $\epsilon >0$. 
Then 
\begin{equation}P(\lvert B \rvert \geq \epsilon) \leq \Delta \max_i P(\lvert A_i \rvert \geq \epsilon/\Delta).
\end{equation}
\begin{proof}
We have 
\begin{align}
\textstyle P(\lvert \sum_{i=1}^\Delta A_i \rvert \geq \epsilon) & \textstyle\leq P( \sum_{i=1}^\Delta \lvert A_i \rvert \geq \epsilon)\\
& \textstyle\leq P( \sum_{i=1}^\Delta \max_i \lvert A_i \rvert \geq \epsilon)\\
& \textstyle= P(\Delta \max_i \lvert A_i \rvert \geq \epsilon).
\end{align}
Using a union bound, one gets \[ \textstyle P(\lvert \sum_{i=1}^\Delta A_i \rvert \geq \epsilon) \leq \Delta \max_i P(\lvert A_i \rvert \geq \epsilon/\Delta ).\]
\label{th:proposition_slow_down}
\end{proof}

Corollary~\ref{th:corollary_slow_down} follows from Proposition~\ref{th:proposition_slow_down}. We assume we have a dependence length of $\tau$. We put the elements of the trajectory into subgroups of $\nu = T/\tau$ independent elements.

We write $B = \sum_{i=1}^\tau A_i$, where $A_i = \sum_{j = 1}^{\nu} s^{[g]}_{(i + j\cdot \tau)}$. Then we apply Proposition~\ref{th:proposition_slow_down} with $\Delta = \tau$ and $\epsilon \leftarrow \epsilon T $. We get 
\begin{align}
P(\lvert \hat \mu_g \rvert \geq \epsilon) &= \textstyle P(\lvert \sum_{i=1}^\tau \sum_{j = 1}^{\nu} s^{[g]}_{(i + j\cdot \tau)} \rvert \geq \epsilon T)\\ 
 &\leq \textstyle  \tau \max_i P(\lvert \sum_{j = 1}^{\nu} s^{[g]}_{(i + j\cdot \tau)} \rvert \geq \epsilon T/\tau ).
 \end{align}
For a given $i$, the $s^{[g]}_{(i + j\cdot \tau)}$ are i.i.d. sub-exponential random variables and hence we can apply Theorem~\ref{th:subexp_decay} and get
\begin{align}
 \textstyle &P\left ( \left\lvert \hat \mu_g \right\rvert  \geq \epsilon  \right) \leq 2 \exp \left( -c \min( K^2/\sigma^2, \epsilon/K)T/\tau \right).
\end{align}
\end{proposition}

\section{Comparison between the median and the mean}
\label{app-compMedianMean}

In order to compare the capability of the STIM when using the mean (STIMean) or the median (STIMedian), we reproduce two figures from the text.

\begin{figure}
  \centering
  { \includegraphics[width=0.4\textwidth]{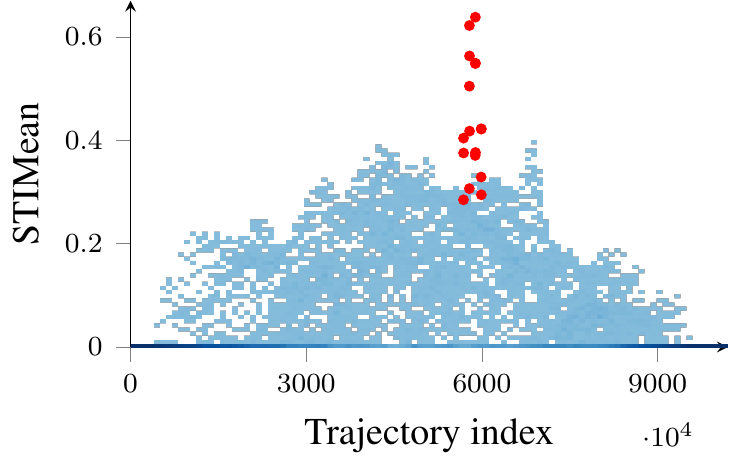}}
  { \includegraphics[width=0.4\textwidth]{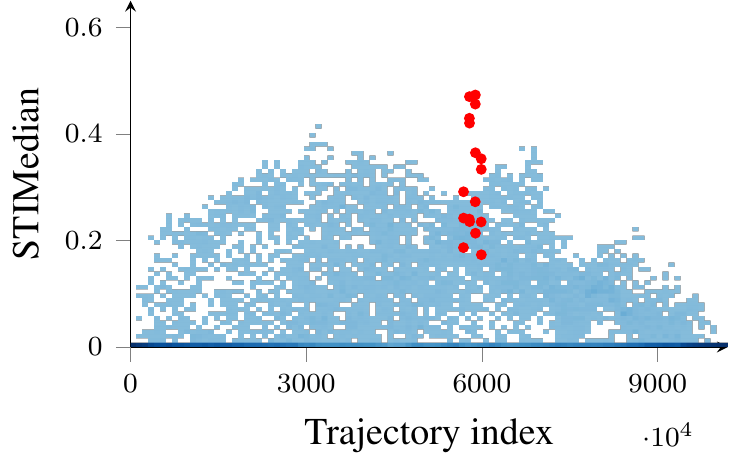}}
  \caption{Reproduction of Figure~\ref{fig:example_plot_mu_sigma_map_sphere} for the mean (top) and for the median (bottom). The planetary signal is found to emerge from the noise more clearly for the mean than for the median.} \label{fig:HD_206893_comparison}
\end{figure}

In Figure~\ref{fig:HD_206893_comparison}, we compare the capability of the STIMean and the STIMedian on HD 20689. We can see that the median performs poorly compared to the mean.

\begin{figure}
  \centering
  { \includegraphics[width=0.4\textwidth]{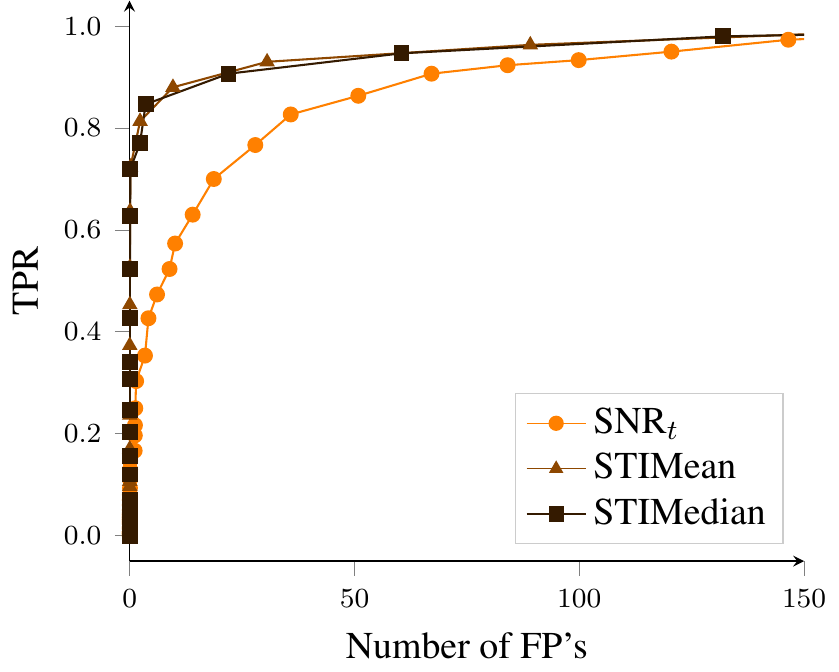}}
  { \includegraphics[width=0.4\textwidth]{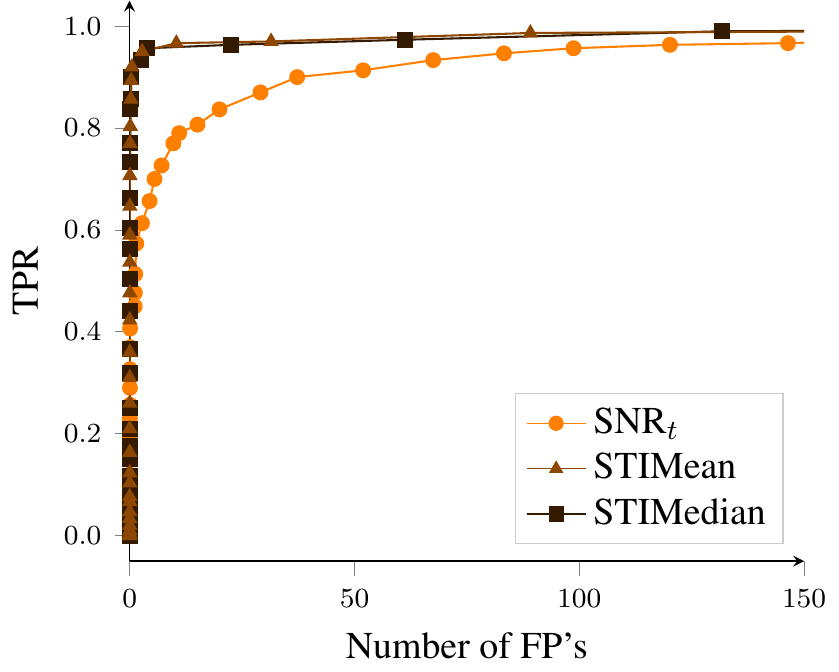}}
  \caption{Reproduction of Figure~\ref{fig:ROC_combined_all_IR_rank.png}, with the STIMedian map added. Top: radial separation of $250 \text{ mas}$ with contrast between  $2.75\times 10^{-5}$ and $4.12\times 10^{-5}$. Bottom: radial separation of $375 \text{ mas}$ with contrast between $1.37\times 10^{-5}$ and $2.75\times 10^{-5}$. We do not observe a significant difference between the use of the median or the mean in the computation of the STIM map.}\label{fig:ROC_comparison}
\end{figure}

We also reproduce the ROC curves of Section with the STIMedian added on Figure~\ref{fig:ROC_comparison}. We do not observe any significant difference between the detection capabilities of the two maps.

\bsp	
\label{lastpage}
\end{document}